\documentclass[article]{JHEP3} 

\usepackage[dvips]{graphicx}
\usepackage{float}
\usepackage{epsfig}
\usepackage{graphicx}
\usepackage{longtable} 
\usepackage{rotating}
\usepackage{amsmath}
\usepackage{amsfonts}
\usepackage{amssymb}
\usepackage{cite}
\usepackage{url}

\newcommand{\sqrts}{\sqrt{s}}
\newcommand{\sqrtsnn}{\sqrt{s_{_{\ensuremath{\it{NN}}}}}}
\def\mean#1{\ensuremath{\left<#1\right>}}
\def\ttt#1{\texttt{\small #1}}
\newcommand{\dd}{{\rm d}}
\def\X{{\rm X}}

\def\cO#1{{{\cal{O}}}\left(#1\right)}

\newcommand{\pp}{{$p\mbox{-}p$}}
\newcommand{\AaAa}{{A--A}}

\newcommand{\PbPb}{{Pb--Pb}}
\newcommand{\ppbar}{{$p\mbox{-}\bar{p}$}}
\newcommand{\aS}{{\alpha_{_S}}}

\newcommand{\pythia}{{\sc pythia}}
\newcommand{\incnlo}{{\sc incnlo}}
\newcommand{\lhapdf}{{\sc lhapdf}}

\newcommand{\dzero}{D$\emptyset$}

\newcommand{\pt}{p_{_{T}}}
\newcommand{\pthat}{\hat{p}_{_{T}}}
\newcommand{\xt}{x_{_{T}}}
\newcommand{\GeVc}{${\rm GeV}/c$}
\newcommand{\GeVcc}{${\rm GeV}/c^2$}

\def\sigmainv{\sigma^{\rm inv}}
\def\sigtev{\sigmainv(1.96\ {\rm TeV}, \xt)}
\def\siglhc{\sigmainv(7\ {\rm TeV}, \xt)}
\def\sigaa{\sigmainv(\sqrts, \xt)}

\title{Single-inclusive production of large-$\pt$ charged particles 
in hadronic collisions at TeV energies and perturbative QCD predictions }

\author{Fran\c{c}ois~Arleo\\
        LAPTH\footnote{Laboratoire d'Annecy-le-Vieux de Physique Th\'eorique, UMR5108}, 
Universit\'e de Savoie, CNRS, BP 110, 74941 Annecy-le-Vieux cedex, France}
\author{David~d'Enterria\\
        ICREA \& Institut Ci\`encies del Cosmos, Univ. Barcelona, 08028 Barcelona, Catalonia}
\author{Andre S. Yoon\\
         Laboratory for Nuclear Science, MIT, Cambridge, MA 02139-4307, USA}

\preprint{\hspace{6.cm} \hepph{1003.2963v2}, {\sf LAPTH-015/10, ICCUB-10-018}}

\abstract{
The single inclusive spectrum of charged particles with transverse momenta
$\pt$~=~3~--~150~\GeVc\ measured at midrapidity by the CDF experiment in proton-antiproton 
(\ppbar) collisions  at $\sqrts$~=~1.96 TeV is compared to next-to-leading order
(NLO) perturbative 
QCD calculations using the most recent parametrizations of the parton distributions 
and parton-to-hadron fragmentation functions. 
Above $\pt\approx$~20~\GeVc, there is a very sizeable disagreement of the Tevatron data
compared to the NLO predictions and to $\xt$-scaling expectations, suggesting a problem 
in the experimental data. 
We also present the predictions for the $\pt$-differential charged hadron spectra 
and the associated theoretical uncertainties for proton-proton (\pp) collisions at 
LHC energies ($\sqrts$~=~0.9~--~14~TeV). Two procedures to estimate the charged hadron spectra 
at LHC heavy-ion collision energies ($\sqrts$~=~2.76, 5.5~TeV) from \pp\ measurements are suggested.}

\keywords{PACS: 12.38.-t 12.38.Bx 13.85.-t 13.87.Fh}

\begin{document}


\section{Introduction}

Hadron production at large transverse momenta ($\pt\gg\Lambda_{\rm QCD}\approx$~0.2~GeV) 
in hadronic interactions originates from the fragmentation of the hard scattered partons produced in the collision.
The presence of a hard scale in the process allows one to employ the powerful theoretical machinery of collinear
factorisation~\cite{Collins:1985ue} to compute the corresponding production cross sections. High-$\pt$ hadron 
cross sections can be thus obtained as a convolution of (i) long-distance universal pieces representing the structure 
of the initial hadrons (parton distribution functions, PDFs) as well as the fragmentation of a final-state quark or gluon 
into the observed hadron (fragmentation functions, FFs), and (ii) short-distance parts that describe the hard partonic 
interactions calculable as a perturbative expansion in terms of the strong running coupling $\aS$. 
The 
measurement of high-$\pt$ hadroproduction in \pp\ and \ppbar\ collisions provides, thus, a valuable 
testing ground of the perturbative regime of Quantum Chromodynamics (pQCD) and of the non-perturbative 
objects (PDFs, FFs) needed to compute a large variety of cross sections at hadronic colliders.

Theoretically, lowest-order (LO) calculations of the inclusive hadron cross sections were performed in the 
late 70s~\cite{Owens:1977sj}, later improved at next-to-leading order (NLO) ~\cite{Aversa:1988vb,ddf,Jager:2002xm} 
and more recently at next-to-leading-log (including soft gluon resummation)~\cite{deFlorian:2005yj,deFlorian:2007ty} 
accuracies. The latest phenomenological developments in this field have focused on 
constraints of the proton PDFs (in particular the polarised ones~\cite{deFlorian:2007ty,Adler:2005in}), 
on improvements of the parton-to-hadron (in particular, gluon-to-hadron) FFs~\cite{Albino:2008gy},
as well as on baseline 
measurements of relevance for high-energy heavy-ions collisions~\cite{d'Enterria:2010zz}.
On the experimental side, 
inclusive unidentified charged hadron production -- i.e. $pp,p\bar{p} \to h^\pm X$, where 
$h^\pm=(h^++h^-)$ is effectively 
the sum of pions (about 60\% of all hadrons), kaons (about 20\% of the total) and protons 
(about 10\% of all hadrons) and their antiparticles -- have been measured above $\pt\approx$~1~\GeVc\ 
at the ISR ($\sqrts$~=~31, 44, 63~GeV)~\cite{ISR_hipt}, at RHIC ($\sqrts$~=~200~GeV)~\cite{RHIC_hipt}, 
Sp$\bar{\rm p}$S ($\sqrts$~=~0.2, 0.5, 0.9~TeV)~\cite{Albajar:1989an}, and Tevatron
($\sqrts$~=~0.63, 1.8, 1.96~TeV)~\cite{Abe:1988yu,Acosta:2001rm,Aaltonen:2009ne} energies. Except at Tevatron, 
the rest of measurements are unfortunately in a moderate $\pt$ range ($\pt\approx$~12~\GeVc\ at most). 
The latest comparisons of the available charged hadron spectra, at RHIC energies~\cite{Adler:2005in,RHIC_hipt},
to NLO calculations~\cite{deFlorian:2005yj} show a good data--theory agreement above 
$\pt\approx$~1.5~\GeVc\ for central and forward rapidities~\cite{d'Enterria:2010zz}. 

In this paper, we compare NLO pQCD calculations to the latest charged particle spectrum measured
at Tevatron and we present predictions with their theoretical uncertainties for the high-$\pt$ hadron 
spectra expected at LHC energies. The motivation is two-fold. 
First, the most recent CDF charged particle spectrum~\cite{Aaltonen:2009ne} covers a very large
$\pt$ range, up to $\pt$~=~150~\GeVc\ where pQCD predictions are reliable and can be confronted to the data. 
Similarly, comparable ``minimum bias'' measurements are expected to be available in the early running of the 
LHC~\cite{lhc_dNdpT}. CMS has already measured a first, yet mostly low-$\pt$, charged hadron spectrum at
$\sqrts$~=~2.36~TeV~\cite{cms2010}. 
Secondly, at the LHC, a \pp\ reference hadron spectrum will be needed at the {\it same} centre-of-mass (c.m.) 
energy as that of heavy-ion (\PbPb) collisions to study  the high-$\pt$ suppression observed
in nucleus-nucleus reactions at RHIC~\cite{wp}. Since the \PbPb\ results will be
nominally obtained at $\sqrtsnn$~=~5.5~TeV a pQCD-based interpolation between the 
results recorded at Tevatron ($\sqrts$~=~1.96~TeV) and during the first LHC \pp\ run 
($\sqrts$~=~7~TeV) 
will be needed.

The paper is organized as follows. In Section~\ref{sec:th} we succinctly remind the theoretical framework 
of our study based on the next-to-leading-order pQCD Monte Carlo (MC) code \incnlo~\cite{INCNLO}. 
In Section~\ref{sec:exp_th}, we compare the charged particle spectra measured at mid-rapidity in \ppbar\ 
collisions at $\sqrts$~=~1.96~TeV~\cite{Aaltonen:2009ne} to the NLO calculations \incnlo\ and to the LO 
parton shower MC \pythia,
as well as to simple $\xt$-scaling expectations. We find that the {\it maximum} theoretical uncertainties 
of the NLO prediction -- associated to the PDF, FF and scale variations added in quadrature -- are $\pm$40\%. 
For increasing transverse momenta, the data is a factor up to 3 orders of magnitude 
larger than the perturbative predictions.
We conclude that above $\pt\approx$~20~\GeVc, there is no possibility to accommodate the data--theory 
discrepancy even accounting for possible additional contributions to the charged particle yield coming e.g. 
from heavy-quarks or vector-boson (plus jet) production. The fact that the parent jet $\pt$-differential 
spectrum is, on the contrary, well reproduced by NLO calculations and that the single particle data also 
violate simple $\xt$-scaling expectations, 
suggest a problem in the experimental results at the highest $\pt$ values.
Finally, in Section~\ref{sec:lhc} we present the charged hadron spectra and associated uncertainties 
predicted by \incnlo\ 
in \pp\ collisions in the range of energies covered by the LHC ($\sqrts$~=~0.9~--~14~TeV), and propose
two methods to determine the \pp\ spectra at intermediate LHC energies of relevance for heavy-ion
running.


\section{Hadroproduction in factorised pQCD}
\label{sec:th}

The inclusive cross section for the production of a single hadron, differential in transverse momentum 
$\pt$ and rapidity $y$, takes the following form at next-to-leading order in $\aS$~\cite{Aurenche:1999nz}:
\begin{eqnarray}
{\dd\sigma\over \dd{\bf \pt}\dd{y}} &=& \sum_{i,j,k=q,g} \int \dd{x_1}\ \dd{x_2}\ 
F_{i/p}(x_1, \mu_{_{F}})\ F_{j/p}(x_2,\mu_{_{F}})\ {\dd{z} \over z^2}\ D_k^h(z,\mu_{_{FF}}) 
\nonumber \\
&&{}\times \left [\left ( {\aS (\mu_{_{R}} ) \over 2 \pi} \right )^2
{\dd\widehat{\sigma}_{ij,k} \over \dd{\bf \pt} \dd{y}}
+ \left ( {\aS(\mu_{_{R}}) \over 2 \pi} \right )^3 K_{ij,k}(\mu_{_{R}} , \mu_{_{F}}, \mu_{_{FF}})
\right ]. 
\label{eq:dsigma_pQCD}
\end{eqnarray}
Here $F_{i/p}(x_1, \mu_{_{F}})$ are the PDFs of the incoming protons $p$ at parton momentum fraction $x$, 
$D_k^h(z, \mu_{_{FF}})$ are the parton-to-hadron FFs describing the transition of the parton $k$ into 
an unidentified hadron $h$ carrying a fraction $z$ of its momentum, $\dd\widehat{\sigma}_{ij,k}/\dd{\bf \pt} \dd{y}$ 
is the Born cross section of the subprocess $i + j \to k + \X$, and $K_{ij,k}$ is the corresponding higher-order term 
(the full kinematic dependence is omitted for clarity). In this paper, we use the \incnlo\ programme~\cite{INCNLO} 
to compute the cross sections, supplemented with various PDFs and FFs sets (see below).
The truncation of the perturbative series at next-to-leading order accuracy in $\aS$ introduces an artificial 
dependence with magnitude $\cO{\aS^3}$, of the cross section on initial-state ($\mu_{_{F}}$) and final-state 
($\mu_{_{FF}}$) factorization scales, as well as on the renormalization scale $\mu_{_{R}}$. 
The choice of scales is arbitrary but the standard procedure is to choose a value around the natural physical scale 
of the hard scattering process, here given by the $\pt$ of the produced hadron. 
We consider below scale variations $\mu_{_{R}},\mu_{_{F}},\mu_{_{FF}}=\kappa\, \pt$, with $\kappa=0.5$~--~$2$
to gauge the theoretical uncertainty linked to the neglected higher-order terms. Hereafter, whenever the scales $\mu_{_R}$, 
$\mu_{_F}$ and $\mu_{_{FF}},$ are given a common value, the latter is denoted $\mu$.

The two non-perturbative inputs of Eq.~(\ref{eq:dsigma_pQCD}) are the parton densities and 
the fragmentation functions. The former are mostly obtained from global-fit analyses of HERA 
proton structure function data, the latter from hadron production results in $e^+e^-$ collisions. 
We use here the three latest PDFs parametrisations available: CTEQ6.6~\cite{cteq66}, 
MSTW08~\cite{mstw08} and NNPDF1.2~\cite{nnpdf12}, included in the \lhapdf\ (version 5.7.1) package~\cite{lhapdf}, 
which take into account the most up-to-date data 
from deep-inelastic lepton-proton scattering and hadronic collisions as well as various theoretical improvements. 
For the 
fragmentation functions into hadrons, we use and compare the three more recent 
FF sets available: DSS~\cite{dss}, AKK08~\cite{akk08} and HKNS~\cite{hkns}, 
which, except for the latter, include for the first time also hadron-hadron collision data 
in their global analyses. These new FF fits cover a larger $z$ range and are more sensitive to the gluon 
fragmentation which dominates high-$\pt$ hadron production in \pp\ collisions~\cite{Albino:2008gy}.

For transverse momenta close to the phase space boundary where the $p_{T}$ of the hadron 
is about half of the partonic centre-of-mass energy ($\xt = 2 \pt/\sqrts \approx 0.1 - 1)$, the 
coefficients of the perturbative expansion are enhanced by extra powers of logarithmic terms of the form 
$\aS^n\ln^{2n-1}(1-\xt)/(1-\xt)$~\cite{resumm}. Resummation to all orders of such 
``threshold'' terms -- which appear when 
the initial partons have just enough energy to produce 
the high-$\pt$ hadron -- have been carried out at next-to-leading logarithmic (NLL) accuracy 
in~\cite{deFlorian:2005yj,deFlorian:2007ty}. Interestingly, the NLL results provide a much reduced 
scale-dependence than the NLO approximation. 
The presently used fixed-order calculations (\incnlo) do not include threshold resummations but 
their effect in the final spectrum is expected to be less important since the typical charged hadron 
$\pt$ range covered by the Tevatron and LHC experiments, $\xt\equiv 2\pt/\sqrts \approx 10^{-4} - 10^{-1}$, 
is far away from the region where such effects start to play a role.


\section{Tevatron data versus perturbative QCD}
\label{sec:exp_th}

In this Section we compare the high-$\pt$ charged particle spectrum measured by the CDF collaboration~\cite{Aaltonen:2009ne}
in the pseudorapidity range $|\eta|~<~1$
to the predictions of \incnlo~\cite{INCNLO} and \pythia~\cite{pythia} MCs and to simple perturbative expectations based on 
$\xt$-scaling~\cite{Brodsky:1973kr}. The measured spectrum covers the range 
$p_T$~=~0.4~--~150~\GeVc, but a comparison to pQCD predictions is only meaningful at high enough $p_T$; 
therefore we impose a minimal cut of $p_T=$~3~\GeVc.
For the NLO analysis, we study separately the effects on the 
spectrum of varying in the calculations the three theoretical scales ($\mu=\pt/2, \pt, 2\pt$), PDFs  (MSTW08, CTEQ6.6
and NNPDF1.2) and FFs (AKK08, DSS and HKNS). We use \pythia\ to determine possible extra contributions 
to the measured high-$\pt$ tracks spectrum coming from heavy-quark fragmentation as well as from real 
and virtual vector-boson 
production, either single-inclusive or in association with a jet. 

\subsection{Data versus \incnlo}
\label{sec:tevatron_incnlo}

\begin{figure}[htbp] 
\centering
\epsfig{width=10.cm,file=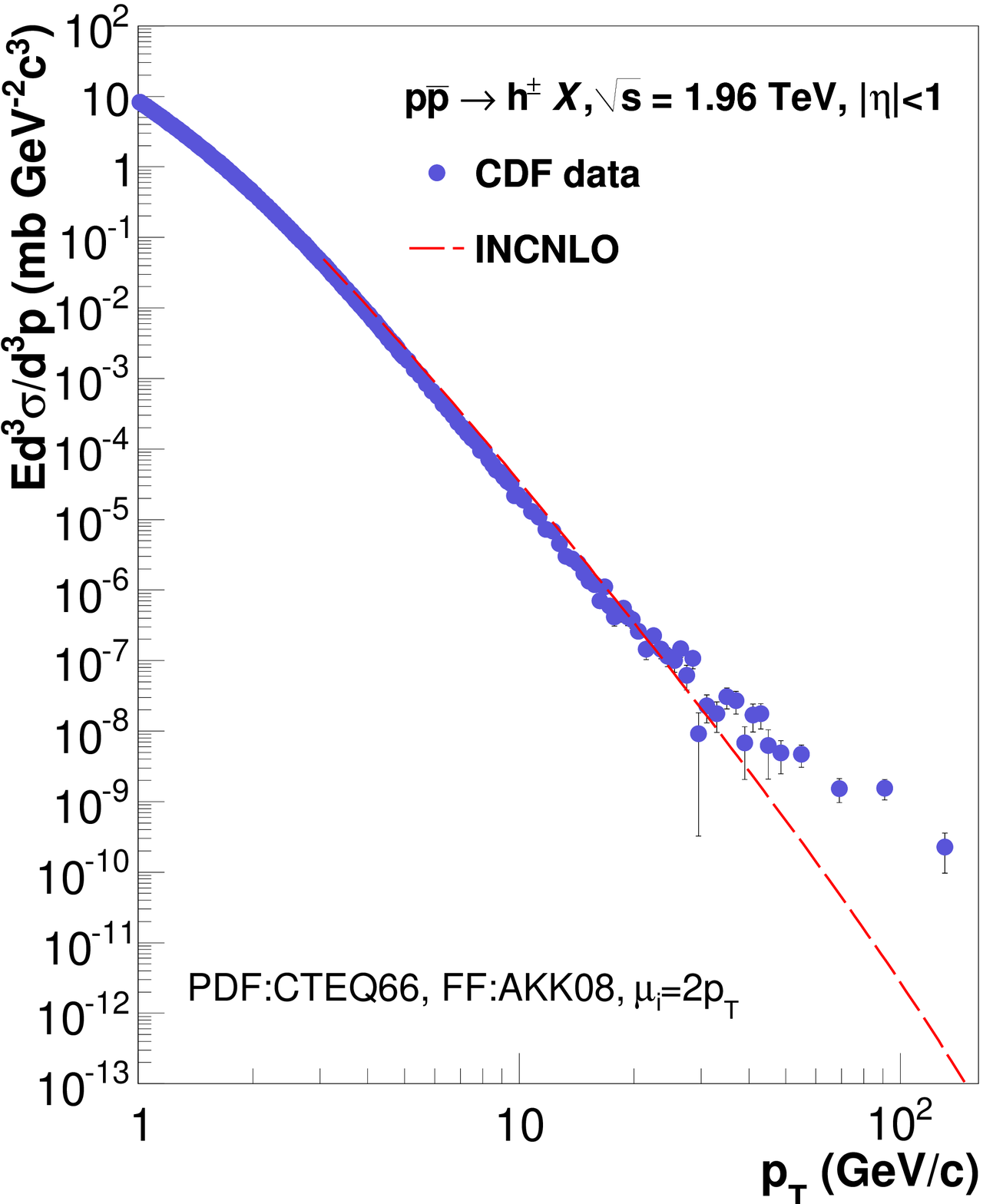}
\epsfig{width=9.8cm,file=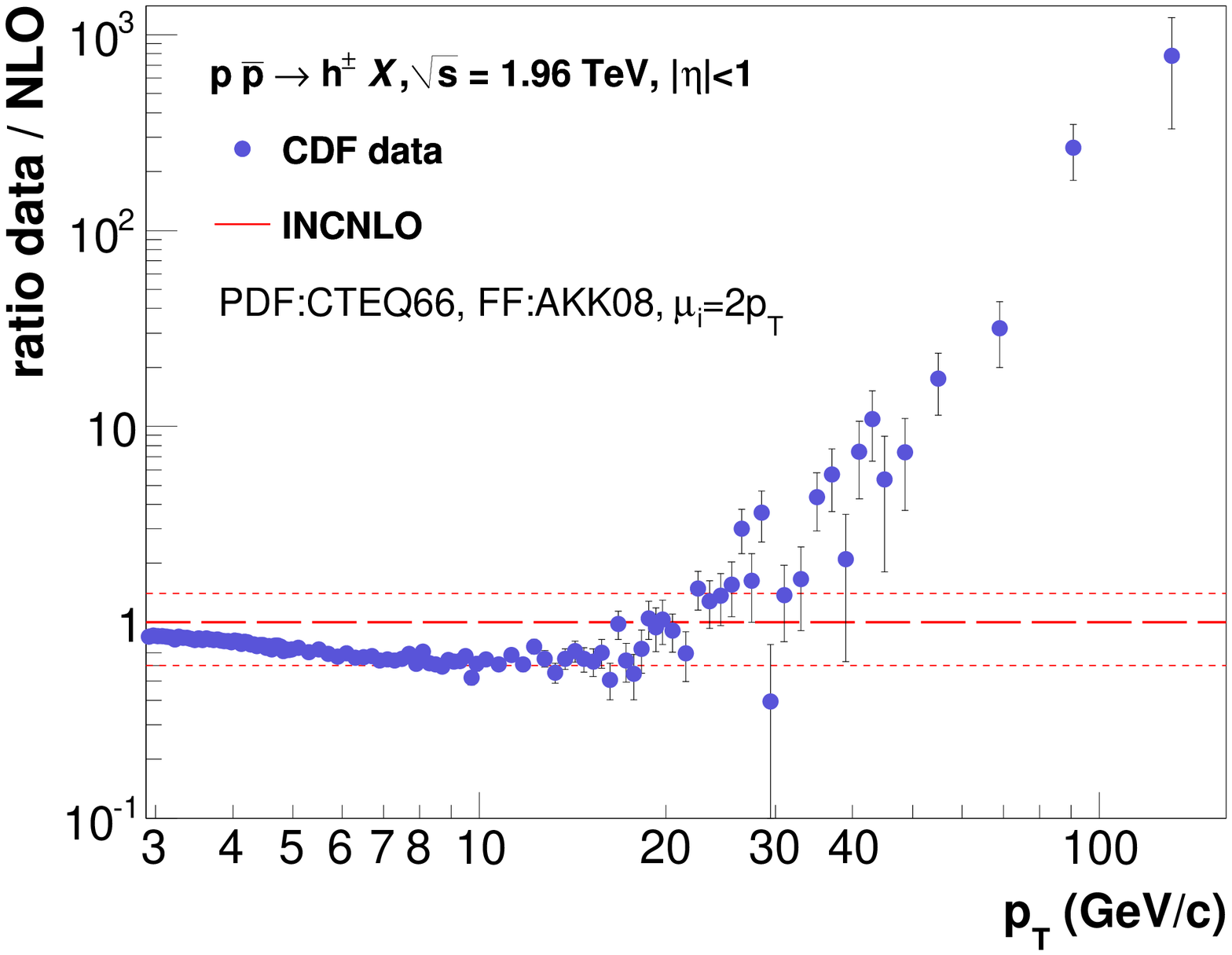}
\caption{Top: Comparison of the charged particle $\pt$ spectrum measured by CDF in \ppbar\ collisions at 
$\sqrts$~=~1.96~TeV~\cite{Aaltonen:2009ne} to NLO pQCD predictions with PDFs fixed to CTEQ6.6, 
scales to $\mu~=~2\pt$, and FFs to AKK08. Bottom: Corresponding ratio of CDF data over theory.
The dashed lines indicate the maximum $\pm$40\% theoretical uncertainty of the calculations (see text).}
\label{fig:data_nlo}
\end{figure}

The measured CDF charged particle \ppbar\ single inclusive distribution is compared to the \incnlo\ predictions 
for charged hadrons in Fig.~\ref{fig:data_nlo}. First, we note that the measured primary track spectrum 
is not corrected for contributions from charged particles other than hadrons. Possible contamination from 
stable leptons (electrons, positrons and muons) are not in principle removed from the measured spectrum. 
As we discuss {\it a posteriori} in Section~\ref{sec:pythia}, those amount however only to a small fraction 
(a few percents) of the total charged particle tracks coming from quark and gluon jet fragmentation 
according to our \pythia\ simulations. 
The \incnlo\ prediction shown in Fig.~\ref{fig:data_nlo} is that which best fits the 
(low $\pt$ range of) the experimental results.
We see that below $\pt\approx$~20~\GeVc\, data and theory agree well for the choice of
scales $\mu=2\pt$, CTEQ6.6 parton densities, and AKK08 parton-to-hadron fragmentation functions. 
Above this $\pt$ value, the CDF spectrum starts to rapidly deviate from the theoretical predictions. 
At the highest transverse momenta the data is up to a factor 800 above the NLO calculations. A very 
conservative quadratic sum of all uncertainties discussed hereafter -- amounting to $\pm$30\% for the 
scales, $\pm$10\% for the PDFs, and $\pm$25\% for the FFs choices -- would result in a maximum theoretical 
uncertainty of $\pm$40\% in the yields (dashed lines around the data/theory ratio).

\begin{figure}[htbp] 
\centering
\epsfig{width=0.66\columnwidth,height=7.cm,file=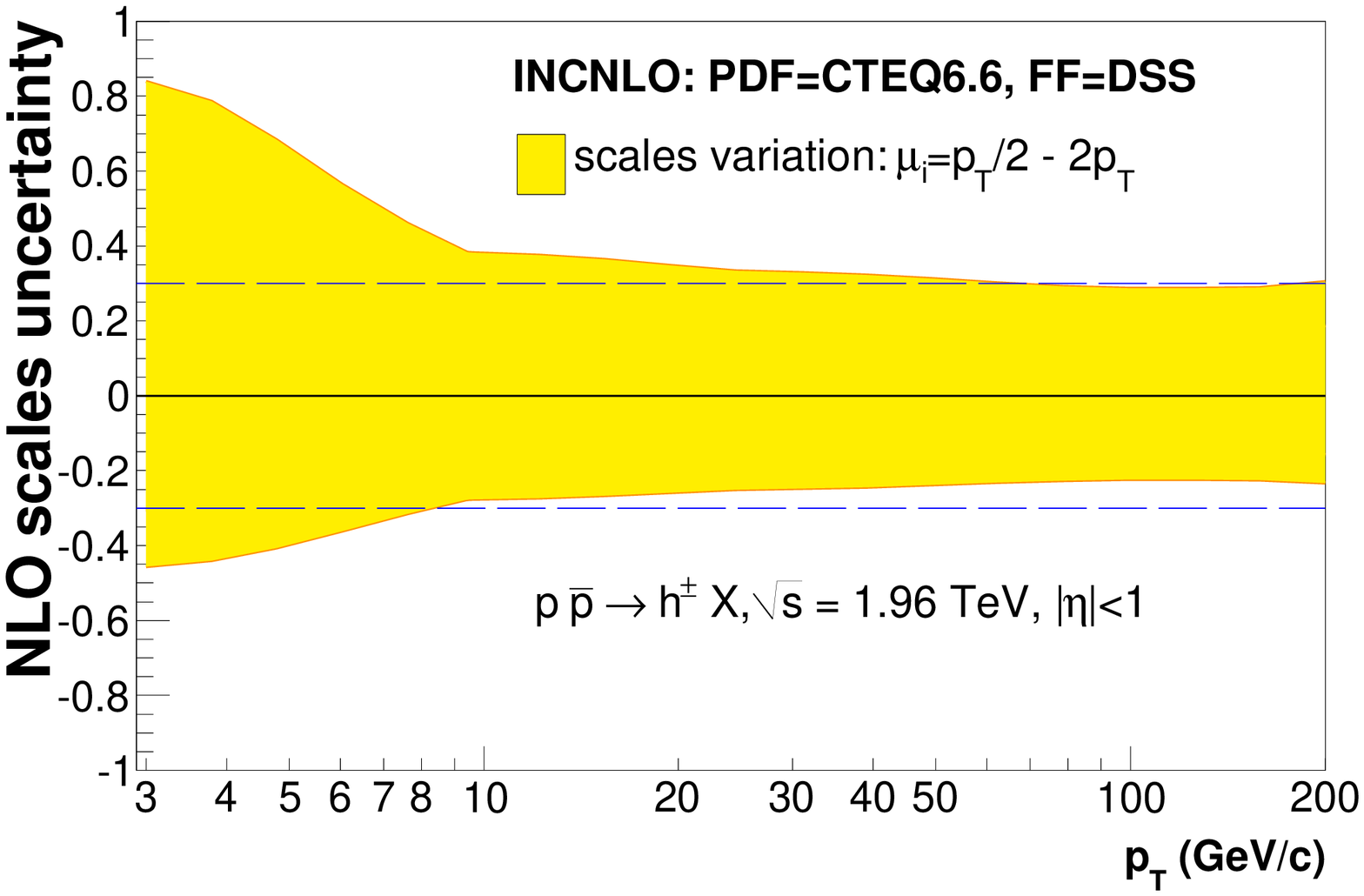}
\epsfig{width=0.66\columnwidth,height=7.cm,file=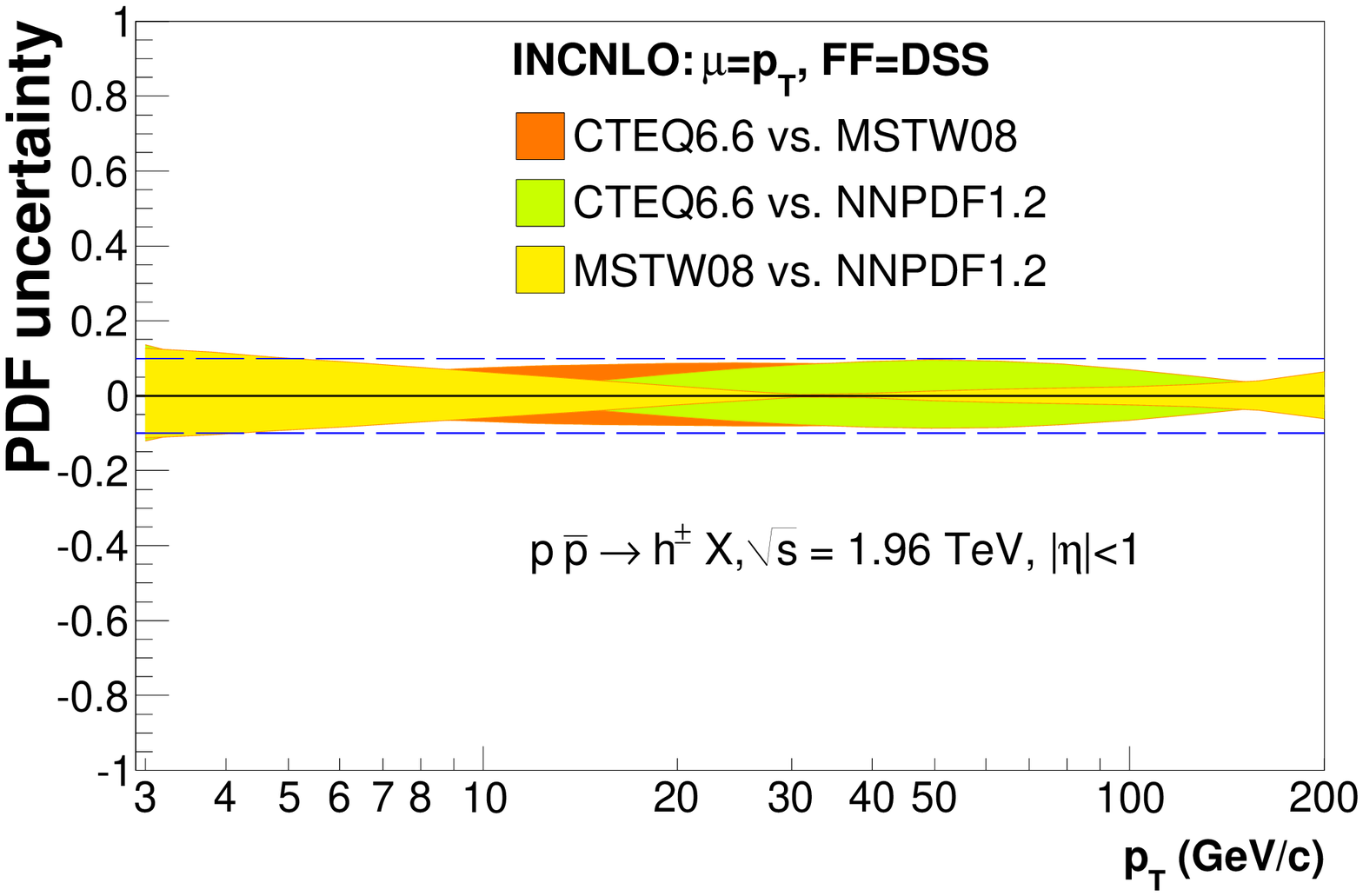}
\epsfig{width=0.66\columnwidth,height=7.cm,file=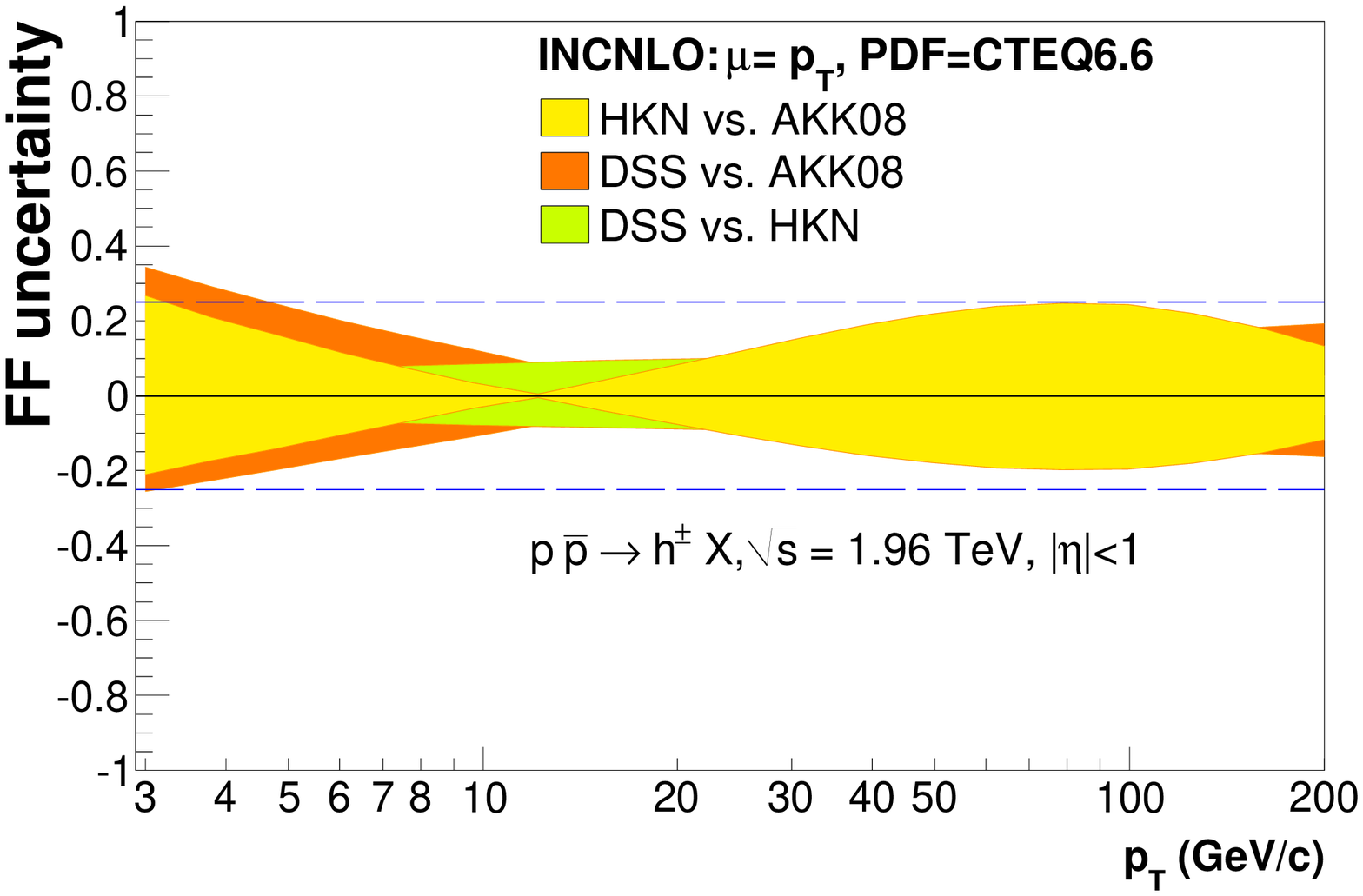}
\caption{Fractional differences between the \incnlo\ charged hadron spectra in \ppbar\ 
at $\sqrts$~=~1.96~TeV for varying scales $\mu_i$, PDF and FF.
Top: Scale uncertainty obtained for fixed PDF (CTEQ6.6) and FF (DSS) varying all three 
scales within $\mu_i~=~\pt/2 - 2\pt$ (the dashed lines indicate a $\pm$30\% uncertainty).
Middle: PDF uncertainty obtained for fixed $\mu=\pt$ and FF (DSS) with three PDFs: 
CTEQ6.6, MSTW08, NNPDF1.2 (the dashed lines indicate $\pm$10\% differences).
Bottom: FF uncertainty obtained for fixed scales ($\mu=\pt$) and PDF (CTEQ6.6) with three 
FFs: AKK08, DSS, HKNS (the dashed lines represent $\pm$25\%).}
\label{fig:nlo_uncertainties}
\end{figure}

\paragraph{\underline{Scale uncertainty}:}
First, in the \incnlo\ calculations we have fixed the PDFs and FFs to 
the CTEQ6.6 and DSS sets respectively\footnote{The choice is in principle arbitrary, other
PDF and FF combinations yield similar results for the scale dependence.}, 
and computed the corresponding spectra setting the three theoretical scales 
to three different values $\mu_{_{F,M,FF}}=\pt/2, \pt, 2\pt$ in all possible 
27 combinations. 
The corresponding range of predictions is shown in Fig.~\ref{fig:nlo_uncertainties} (top) where we plot
a shaded band covering the whole range of fractional differences between the spectra obtained for any
choice of scales. The ``closest-to-the-average'' spectrum is obtained setting all scales to $\mu=\pt$.
The largest (resp. lowest) charged hadron yield predictions are obtained with mostly all three scale 
values set to $\mu_i=\pt/2$ (resp. $\mu_i=2\pt$).
At low $\pt$ the scale uncertainty is quite large (indicating as expected larger higher-order corrections) 
but otherwise above $\pt$~=~10~\GeVc\ it stays roughly 
constant at around $\pm$30\% up to the highest momenta considered (dashed lines in the figure).

\paragraph{\underline{PDF uncertainty}:}

Second, in the middle panel of Fig.~\ref{fig:nlo_uncertainties} we show as a function of $\pt$ 
the theoretical uncertainty 
associated to the PDF choice. It has been obtained with \incnlo\ comparing the fractional 
differences between the single charged hadron spectra at $\sqrts$~=~1.96 TeV for fixed scales 
($\mu = \pt$) and FF (DSS) and three different PDFs. The dashed bands 
plotted cover the range of maximum relative differences in the theoretical spectra obtained with MSTW08, 
CTEQ6.6 and NNPDF1.2. Those differences are small, below 10\% and mostly $\pt$-independent (dashed lines).

\paragraph{\underline{FF uncertainty}:}

Last, we have used \incnlo\ complemented with the three latest FFs available in the market: AKK08, DSS 
and HKNS, to compute the charged hadron spectrum for the CDF kinematics, with scales ($\mu~=~\pt$) 
and PDFs (CTEQ6.6) fixed. The main differences between FF sets concern the fractional $\pi^\pm$, $K^\pm$ and $p/\bar{p}$ 
compositions as a function of $\pt$. 
Yet, the total\footnote{As a  cross check, for all FF sets we have confirmed that 
the NLO spectrum obtained from the sum of the spectra individually obtained with the pions, kaons and protons 
FFs is indeed equal to the one obtained with the non-identified charged hadron FFs.} hadron yield predicted 
by the three FFs for \ppbar\ at 1.96~TeV is quite similar as can be seen in the bottom panel of 
Fig.~\ref{fig:nlo_uncertainties} where we plot the relative differences between the spectra computed for varying FFs.
The maximum theoretical uncertainty linked to the FF choice amounts to about $\pm$25\% of the charged hadron
yield at any $\pt$ (dashed lines), although the differences between FFs appear to be smaller at intermediate hadron 
$\pt\approx$~10~--~25~\GeVc.

A conservative quadratic sum of the fractional uncertainties linked to the NLO scales, PDFs and FFs
choices results in a total theoretical uncertainty of order $\pm$40\% whereas the maximum difference
between the data and the calculations amounts to much larger factors, up to $\mathcal{O}(10^3)$ 
at $\pt\gtrsim100$~\GeVc\ (Fig.~\ref{fig:data_nlo}).


\subsection{Data versus \pythia}
\label{sec:pythia}

Given the large discrepancy between the experimental and NLO predictions for the charged particle
spectrum at high-$p_T$ one may wonder whether other charged particles -- apart from 
$\pi^\pm,K^\pm,p/\bar{p}$ coming from the fragmentation of quarks and gluons -- may
contribute in any way to the experimentally measured distribution beyond $\pt\approx$~20~\GeVc. 
A first possibility that we have considered is whether other charged products from charm and bottom 
jets (with relative increasing importance at large transverse momentum) play any role. 
Although the inclusive hadron FFs used in our NLO calculations contain {\it all} pion, kaons and 
(anti)protons issuing from light- as well as heavy-quark fragmentation, charm and bottom 
hadrons decay also into charged leptons\footnote{As a cross check, we have confirmed that the
\pythia\ spectrum of single leptons from $c$ and $b$ production agrees relatively well (within
a factor of two) with more involved fixed-order NLL calculations~\cite{fonll}.} which are not 
included in \incnlo. 
Thus as a independent theoretical check, 
we have computed the inclusive yield of {\it all} charged particles with the \pythia\ MC 
(v6.420)~\cite{pythia} with the D6T\cite{tuneD6T} tuning\footnote{Tune D6T uses the CTEQ6LL PDF and 
describes the underlying event and the Drell-Yan data at Tevatron.}
%
in the ``minimum bias'' and QCD-jets modes (\ttt{MSEL = 1} with low-$\pt$ production, \ttt{ISUB = 95}, 
switched on to correctly simulate the low-$\pt$ region). 
The chosen processes produce not only light-quarks and gluons but also heavy-quarks\footnote{\pythia\ settings: 
\ttt{PARP(91)=2.1} \GeVc\ (intrinsic $k_T$), \ttt{PMAS(4,1)=1.5}~\GeVcc\ ($m_c$ mass), \ttt{PMAS(5,1)=4.8}~\GeVcc\ 
($m_b$ mass), \ttt{MSTP(33)=1} ($K$-factor). Alternative running of {\it standalone} heavy-quark production 
(with \ttt{MSEL = 4} and \ttt{5}) 
would require $K$-factors of 2~--~4 in order to reproduce the heavy flavour $\pt$ spectra measured at various 
colliders~\cite{Lourenco:2006vw}.} including flavour excitation, $Qg\to Qg$, and gluon splitting, $g\to Q\bar{Q}$.


\begin{figure}[htbp]
\centering
\epsfig{width=0.66\columnwidth,height=11.5cm,file=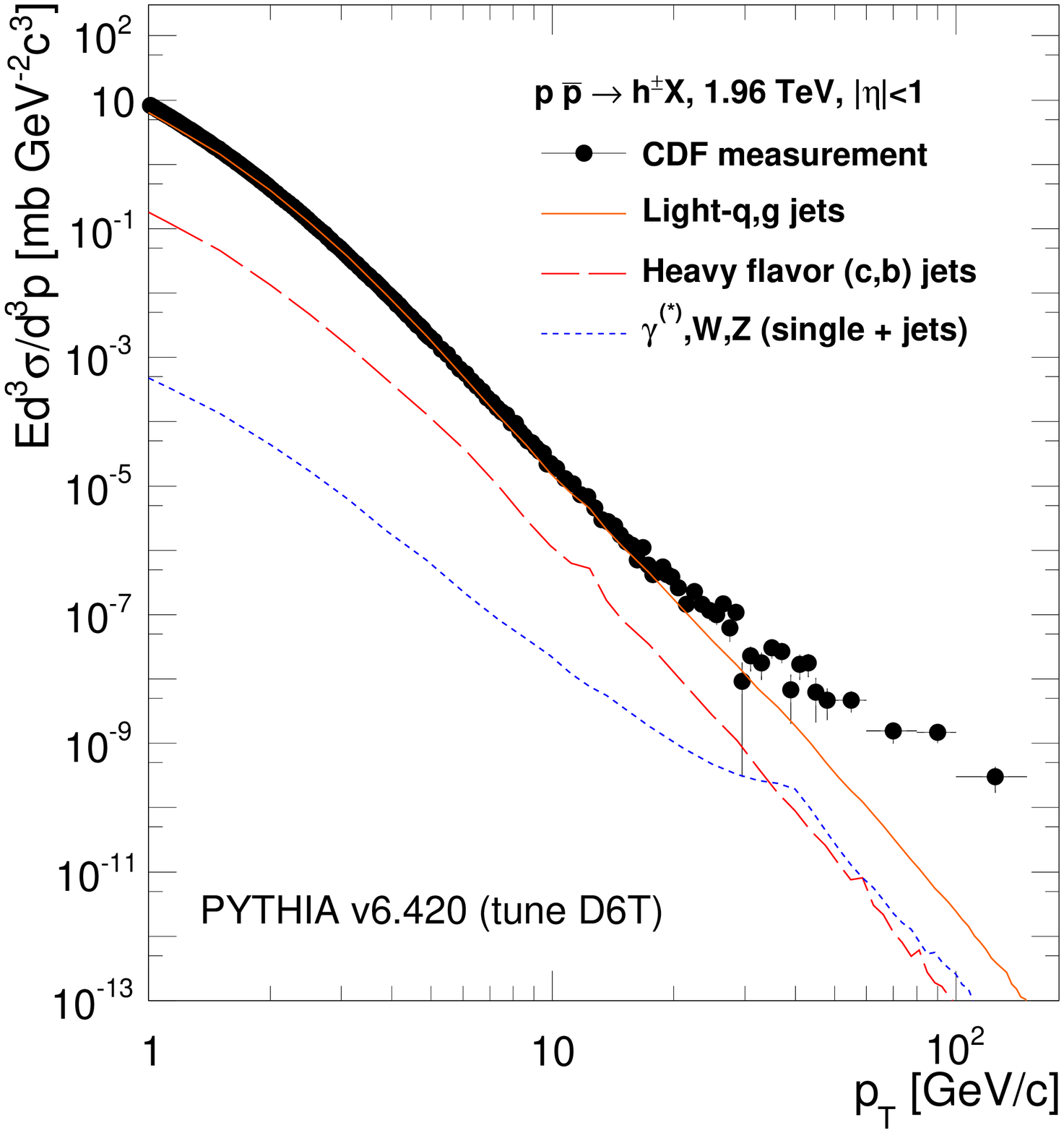}
\epsfig{width=0.66\columnwidth,file=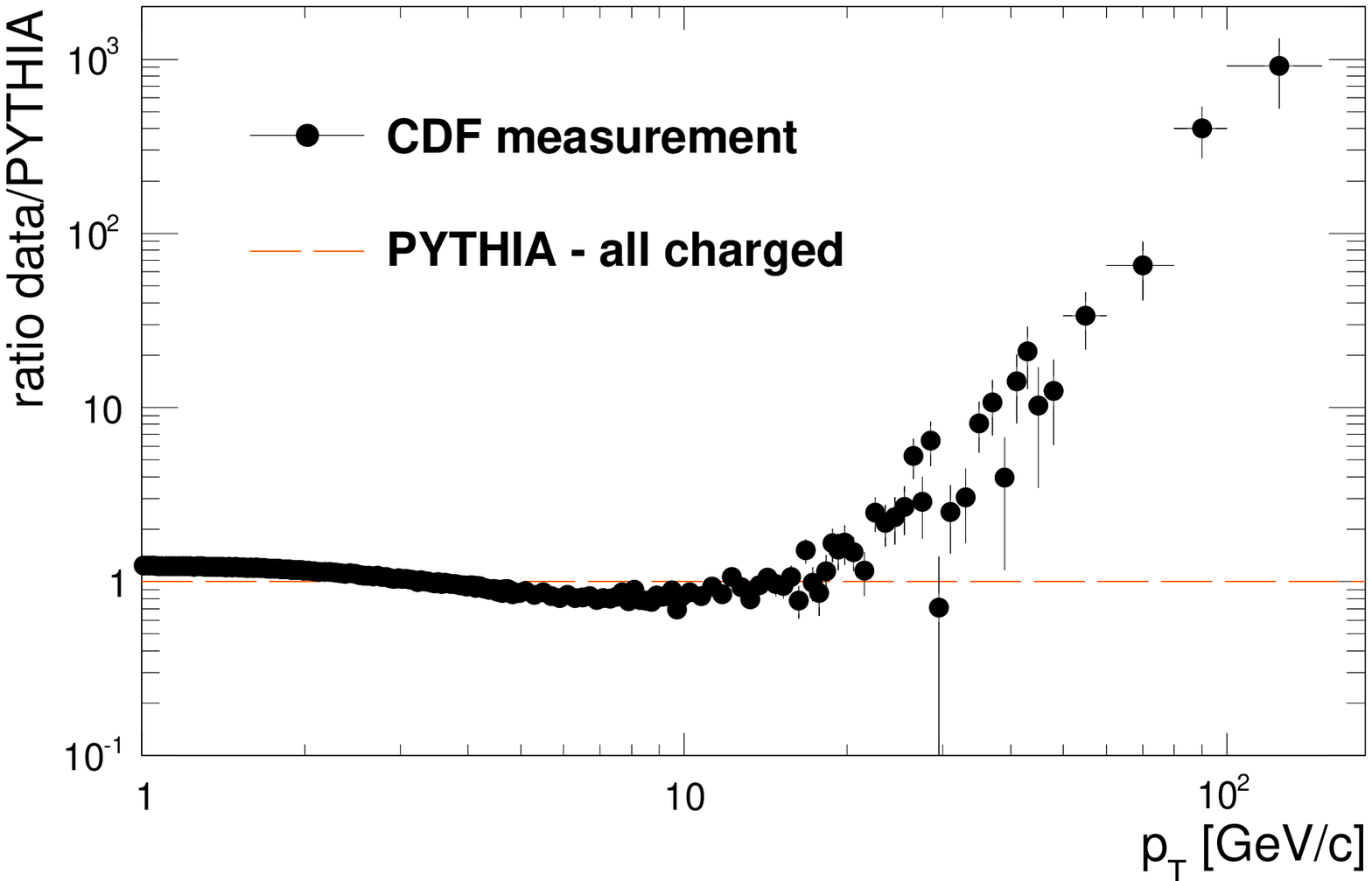}
\caption{Top: Comparison of the CDF data (symbols) to \pythia\  (v6.420, D6T tuning) 
$\pt$ distribution of charged particles in \ppbar\ collisions at $\sqrts$~=~1.96~TeV coming 
from the fragmentation of (i) 
light-quarks and gluons and (ii) heavy-quarks, and (iii) from processes involving the 
production of vector bosons ($\gamma^{(\star)}$, $W^\pm$, $Z^0$).
Bottom: Ratio of the CDF data to the sum of all \pythia\ charged particle contributions.}
\label{fig:pythia_cdf}
\end{figure}

To obtain enough statistics at high-$\pt$, we have run with up to 12 different ranges for the 
minimum and maximum parton momenta in the 2$\to$2 scatterings ($\pthat$~=~0~--~10,~10~--~15, 15~--~20, 
20~--~50~\GeVc,\dots and $\pthat >$~470~\GeVc) weighted by their corresponding cross sections. 
We have then explicitly separated the contributions coming from the fragmentation of high-$\pt$ light-flavours 
($u,d,s$ and gluon) from those coming from charm and bottom quarks. 
As done in CDF, we take all charged particles\footnote{\pythia\ settings: \ttt{MSTJ(22)=2, PARJ(71)=10}.} 
exactly as defined in their analysis (i.e. all primary particles with mean lifetimes 
$\tau~>$~0.3 10$^{-10}$~s and the decay products of those particles with shorter $\tau$).
The results of our studies are shown 
in Fig.~\ref{fig:pythia_cdf}.
The inclusive charged products of $c$-quark and $b$-quark fragmentation represent a very small 
(less than 5\%) fraction of the total yield of particles measured at high-$\pt$ by CDF.

A second possibility that we have explored is whether the charged products of real and virtual 
vector-boson ($\gamma^{(\star)}$, $W^\pm$, and $Z^0$) production 
-- either single-inclusive or in association with a jet --
which start to play a role at increasing transverse momenta, could partially account for the data--theory 
discrepancy. We have run prompt photon production in \pythia\ including the Born-level $\gamma$-jet 
Compton and annihilation diagrams (\ttt{ISUB = 29,14} respectively).
The $W^\pm$, $Z^0$ and DY production (\ttt{ISUB = 1,2,15,16,30,31}),
includes single-inclusive ($2 \to 1$) as well as double-inclusive 
($2 \to 2$) $W^\pm$-,$Z^0$-,DY-jet channels with $\hat{p_T} >$~20~\GeVc. 
The $W^\pm$ and $Z^0$ contributions produce a (local) Jacobian peak in the charged-particle $\pt$ distribution 
at about half the vector-boson mass, $\pt\approx$~40~\GeVc. All those contributions, 
shown added up in Fig.~\ref{fig:pythia_cdf}, increase the charged particle yield by up to 10\% 
in the range above $\pt\approx$~40~\GeVc. This number is consistent with a simple order 
of magnitude estimate based on the ratio of electroweak and strong coupling constants 
valid when $\pt\gg M_W/2$: $\left(\alpha_{_{\rm EW}}/\alpha_s\right)^2=\left(0.034/0.12\right)^2=\cO{10^{-1}}$. 
Clearly, those processes contribute little to the total yield of charged particles and 
therefore cannot justify the observed large discrepancy between data and theory.


\subsection{Data versus $\xt$-scaling}
\label{sec:xtscaling_tevatron}

A robust pQCD prediction for hard processes $A\ B\to C\ X$ in hadronic collisions is the power-law 
scaling of the inclusive invariant cross section, 
\begin{equation}
E\ d^3\sigma/d^3 p = F(\xt)/ \pt^{n(\xt,\sqrts)} = F^\prime(\xt)/\sqrts^{n(\xt,\sqrts)}\;.
\label{eq:scaling}
\end{equation}
In the original parton model 
the power-law fall-off of the spectrum is simply  $n=4$ since the underlying $2 \to 2$ subprocess amplitude for point-like 
partons is scale invariant. In QCD, small scaling violations appear due to the running of $\aS$ and the evolution of PDFs and FFs.
At midrapidity and at fixed $\pt=10$~\GeVc, the power-law exponent computed at NLO accuracy 
increases slowly from $n\simeq$~5 
at small values of $\xt$ ($\xt = 10^{-2}$) up to $n\simeq$~6 at $\xt=0.5$, with a very small 
dependence on the specific hadron species~\cite{Arleo:2009ch}.


\begin{figure}[htbp]
\centering
\epsfig{height=8.4cm,file=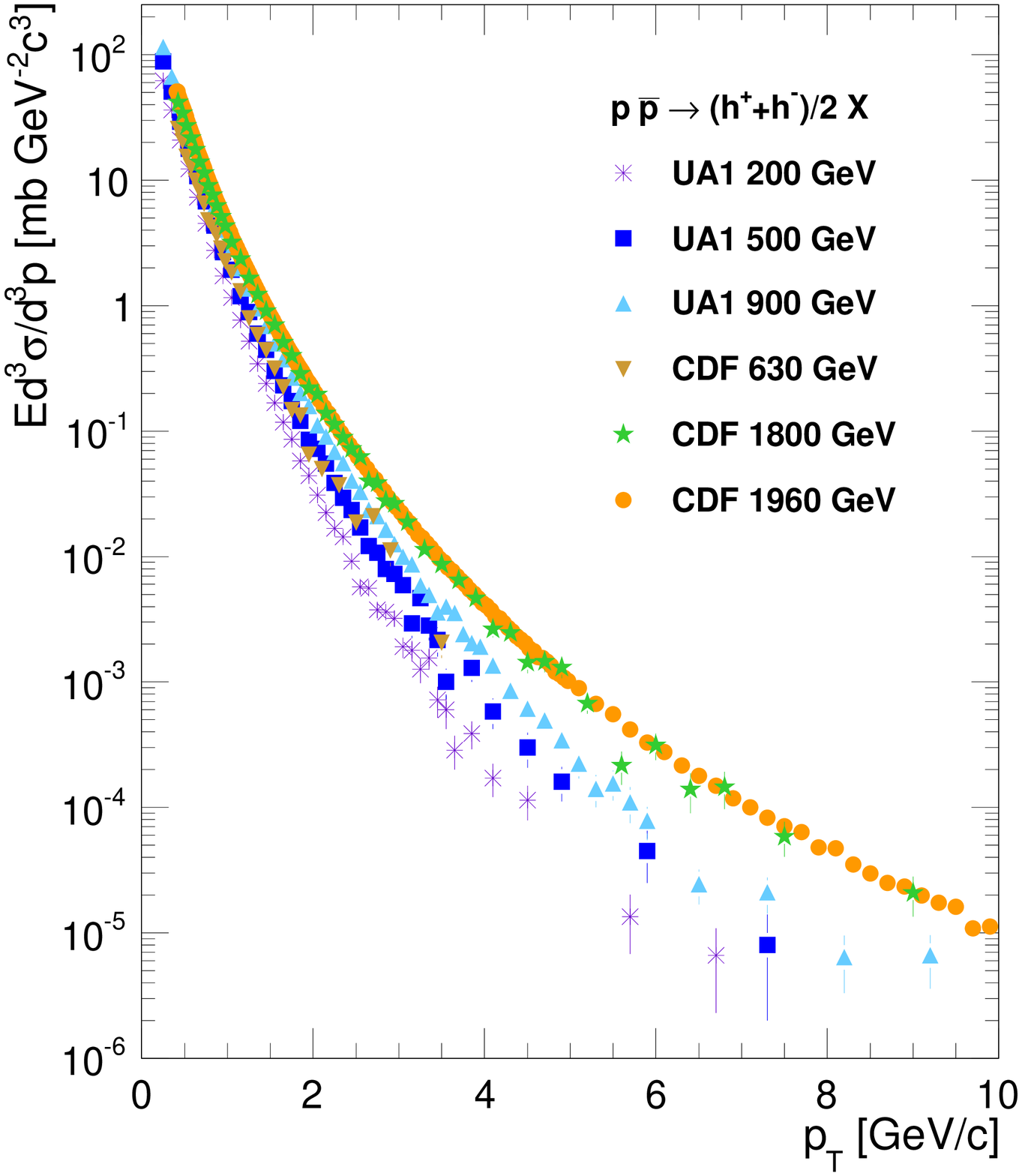}\hfill
\epsfig{height=8.4cm,file=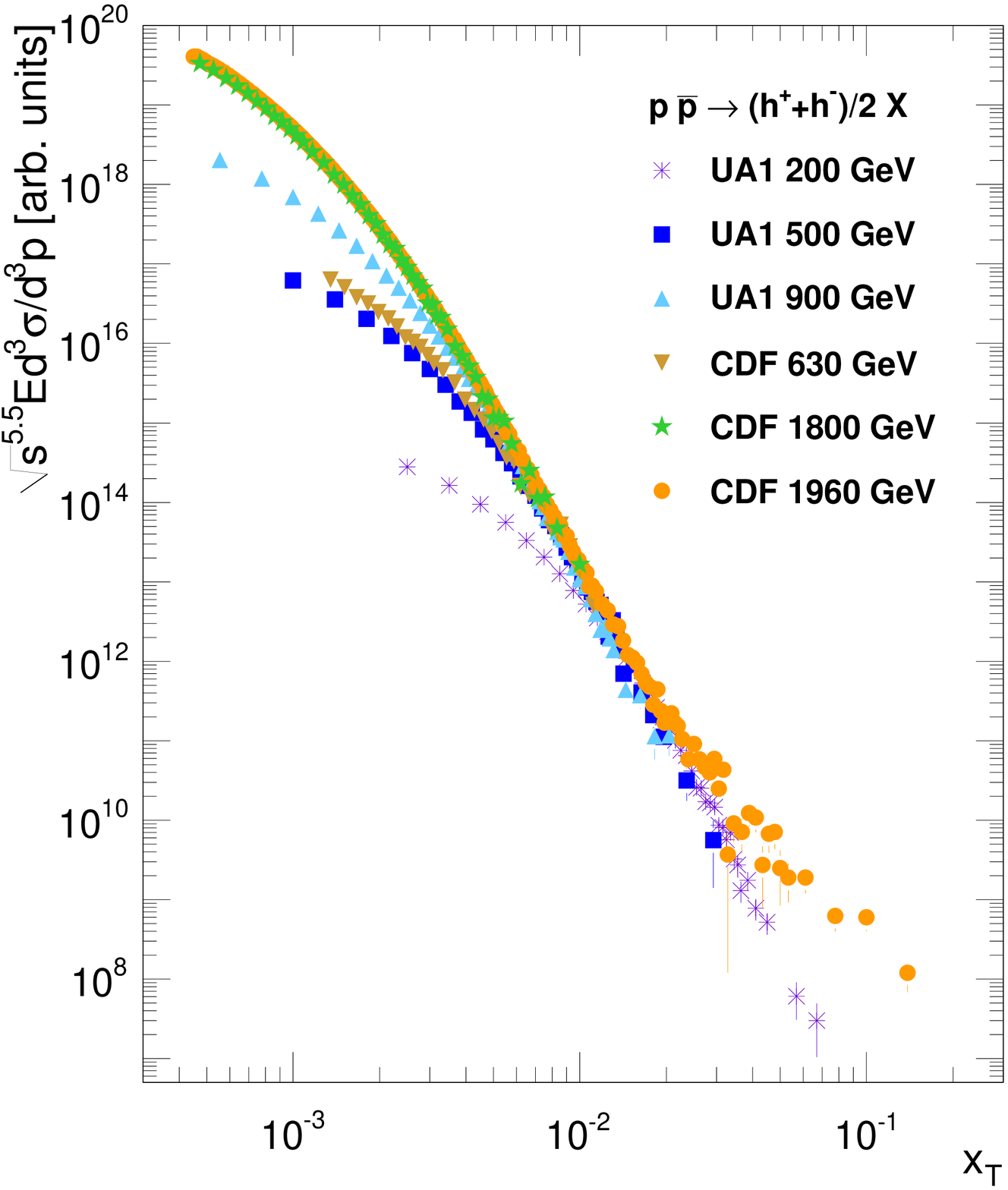}
\caption{Compiled charged particle cross-sections measured in \ppbar\ collisions
at five different c.m. energies from 0.2 TeV to 1.96 TeV 
plotted as a function of $\pt$ (left) and as a function of $\xt$ (right) 
scaled with an effective common exponent of $n$~=~5.5 (see text).}
\label{fig:xT_compiled}
\end{figure}

Except for the latest CDF data, the theoretical expectation, Eq.~(\ref{eq:scaling}), is indeed well 
fulfilled by the experimental charged particle spectra measured so far in \ppbar\ collisions\footnote{A 
factor of 1/2 is applied hereafter to the CDF Run II spectrum as they measured $(h^{+}+h^{-})$ 
instead of the average $0.5\times(h^{+}+h^{-})$ for all other measurements.} at different 
centre-of-mass energies at the CERN Sp$\bar{\rm p}$S ($\sqrts$~=~0.2, 0.5, 0.9~TeV)~\cite{Albajar:1989an} 
and Tevatron ($\sqrts$~=~0.63, 1.8, 1.96~TeV)~\cite{Abe:1988yu,Aaltonen:2009ne} colliders.
All the $\pt$ spectra feature power-law behaviours above $\pt\approx$~2~\GeVc\
(the higher the c.m. energy the smaller the exponent of the fall-off, see Fig.~\ref{fig:xT_compiled} left). 
Following the expectation~Eq.~(\ref{eq:scaling}), in order 
to extract a common $n$ value from these different data sets, 
the measurements are plotted as a function of $\xt$, multiplied by $\sqrts^n$ and fitted
with the following 3-parameter functional form 
\begin{equation}
\sqrts^{n} \frac{E \dd^3\sigma}{\dd^3p}\Bigr\vert_{y=0} =\ p_0\cdot [ 1 + (x_{T}/p_1)]^{p_2} \;.
\label{eq:simple_fit}
\end{equation}
In the data fitting, a minimum $p_T$ of 2~\GeVc\ is applied to exclude the region where 
soft particle production (which does not follow $\xt$-scaling) is dominant, which is consistent 
with what is used in~\cite{Arleo:2009ch}. We also exclude the CDF Run-II data from the global fit 
since, as we see {\it a posteriori}, there is no possibility to get an agreement with the lower energy measurements.
With the obtained $\{p_i\}$-parameters, using $n_{_{\rm NLO}}\approx$~5 as a guidance, 
the exponent $n$ is varied from 4 to 7 in incremental steps in order to minimize the following 
$\chi^{2}$ function with {\sc minuit}~\cite{minuit} 
\begin{equation}
\chi^{2}(n,\{p_i\}) = \sum^{n_{dat}}_{j=1} \Bigg[\frac{\frac{E \dd^3\sigma}{\dd^3p}\Bigr\vert_{y=0}-\left(\frac{\ p_0\cdot [ 1 + (x_{T}/p_1)]^{p_2}}{\sqrts^{n}}\right)} {\sigma_{j}} \Bigg]^{2}\;,
\label{eq:chisquare_min}
\end{equation}
where $\sigma_{j}$ are the quadratic sum of the statistical and systematic experimental uncertainties.
In Fig.~\ref{fig:xT_compiled} (right) we show the experimental charged particle spectra 
scaled by $\sqrts^n$ as a function of $\xt$ with the best value of $n$ obtained from the fit, $n=5.5$.
We note that all measurements spanning a range of one order-of-magnitude in centre-of-mass energies 
follow a universal curve after rescaling up to the highest $\xt\sim$~0.03 measured at lower energy. 
A deviation of the CDF measurement at $\sqrts$~=~1.96~TeV from the trend established by the lower $\sqrts$ 
measurements is prominent above $\xt~\sim$~0.03 ($\pt~\sim$~30~\GeVc).

\begin{figure}[htbp]
\centering
\epsfig{width=0.66\columnwidth,height=11.5cm,file=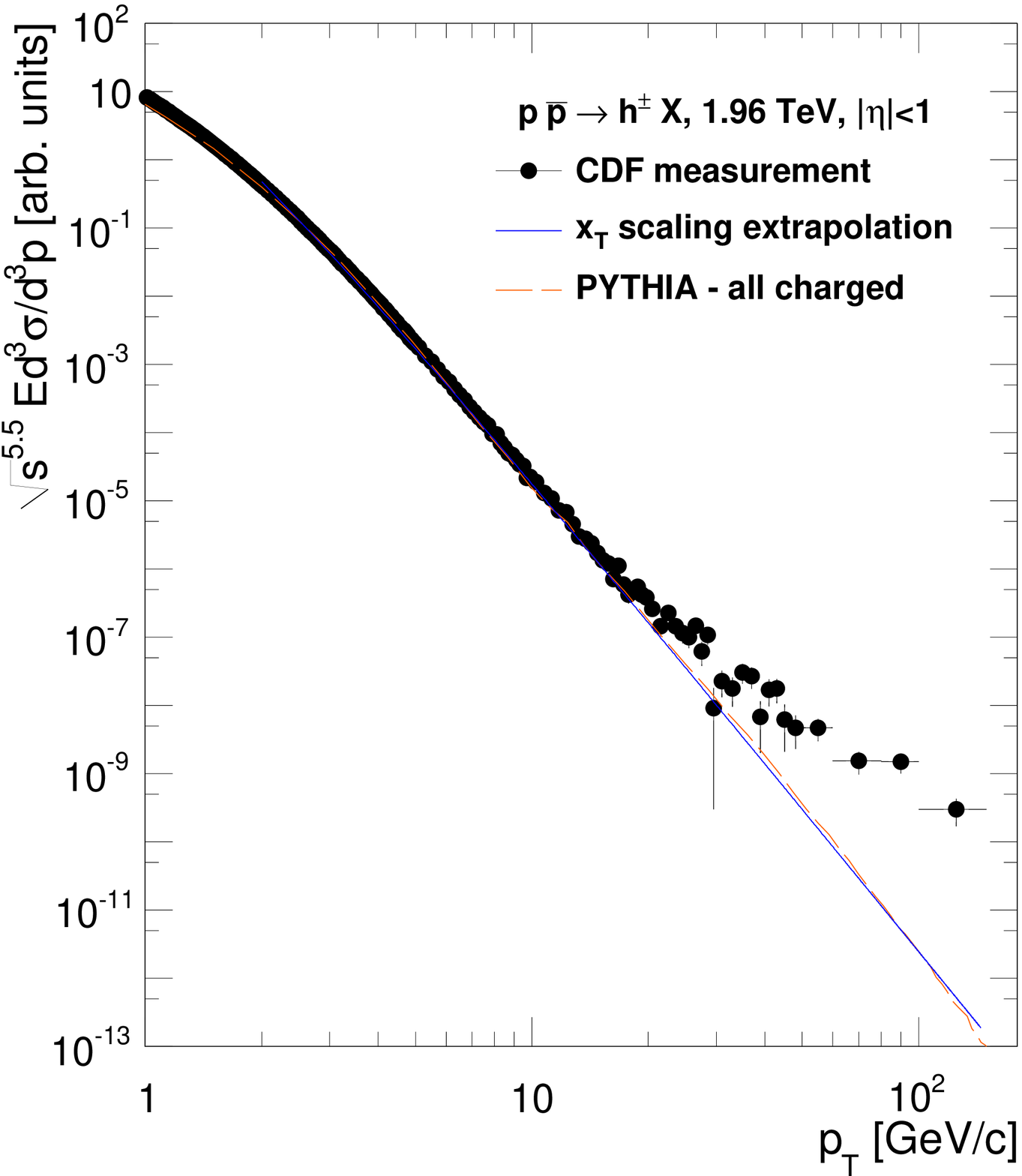}
\epsfig{width=0.66\columnwidth,file=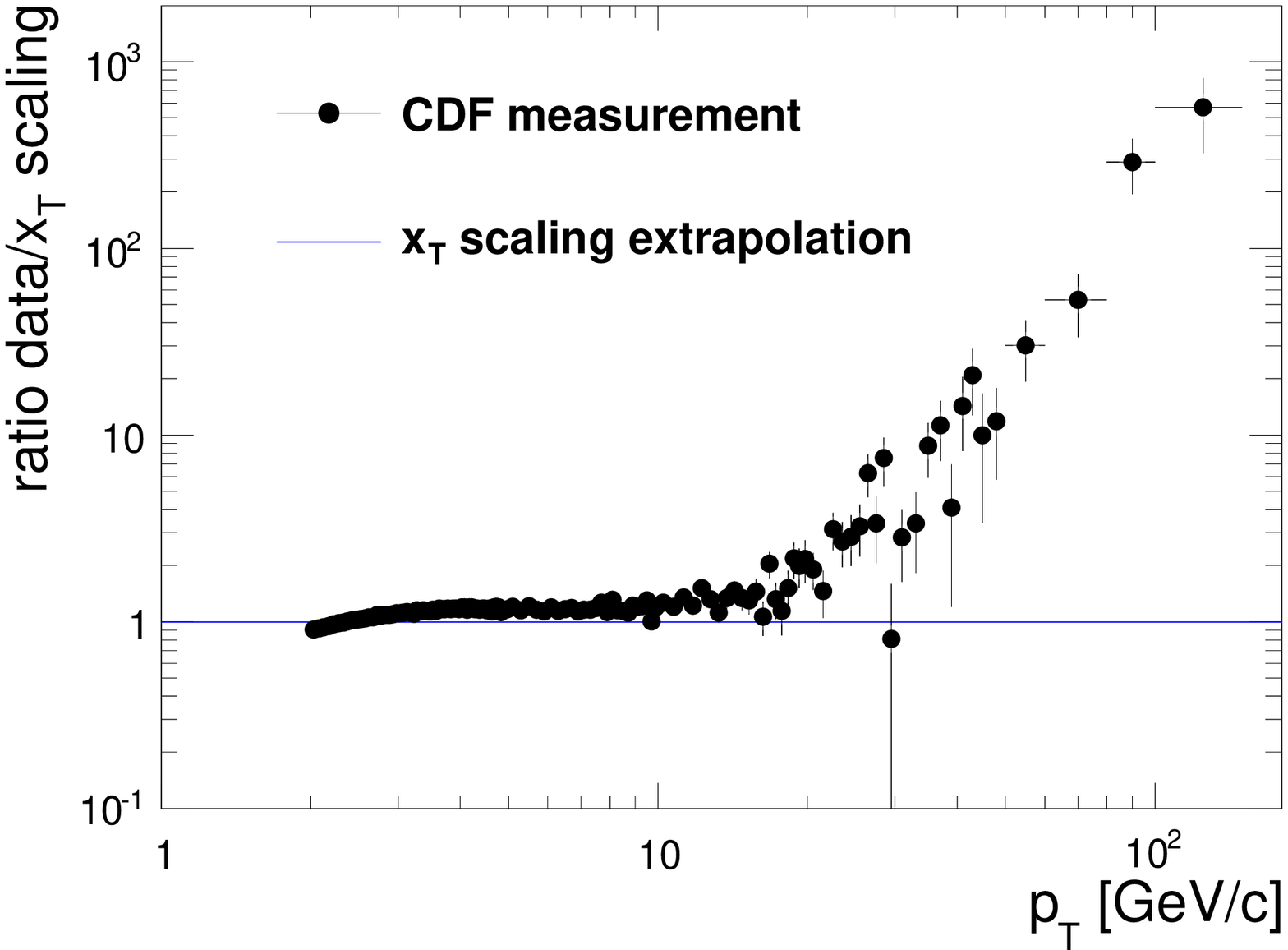}
\caption{Top: Inclusive CDF Run-II charged hadron spectrum (filled circles) compared to the 
$x_T$-scaled extrapolation of lower energy \ppbar\ data (blue solid line), as well as to 
the total \pythia\ prediction (dashed red line) of Fig.~\protect\ref{fig:pythia_cdf}.
Bottom: Corresponding ratio of CDF data over $\xt$-scaling expectation.}
\label{fig:xT_scaling}
\end{figure}

In order to better assess the amount of deviation of the CDF data to the $\xt$-scaling expectation
we show in Fig.~\ref{fig:xT_scaling} (top) the $x_T$-scaled fit obtained from all lower-energy data
extrapolated to an expected $\pt$ spectrum at $\sqrts$~=~1.96~TeV
compared with the CDF Run II measurement and with the \pythia\ prediction shown in 
Fig.~\ref{fig:pythia_cdf}. 
As observed before, beyond $\pt\approx$~20~\GeVc\ the latest CDF data clearly fail to follow
the $\xt$-scaling expectation fulfilled by the rest of charged hadron measurements.

\subsection{Discussion}

The fact that no combination of PDF, FF and/or theoretical scales in the NLO calculations is able to 
reproduce the Tevatron experimental data above $\pt\approx$~20~\GeVc\ by such a large factor is totally unexpected. 
Indeed, similar calculations based on {\sc fastnlo/nlojet++}~\cite{fastnlo,Nagy:2003tz}, 
reproduce perfectly well the inclusive jet spectrum measured in \ppbar\ collisions at $\sqrts$~=~1.96 TeV 
in the range $p_T^{jet}\approx$~50~--~600~\GeVc. The shape and magnitude of the CDF jet 
measurement is well reproduced using CTEQ6.1M PDFs and renormalization and factorization 
scales\footnote{Note that in the inclusive jet calculation there is one scale less, the 
fragmentation one $\mu_{_{F}}$.} set to $\mu_{_{R}} = \mu_{_{F}} = \pt^{jet}/2$~\cite{Aaltonen:2008eq}.
Likewise, the \dzero\ jet measurement agrees well with the same NLO predictions with CTEQ6.5M parton
densities and $\mu_{_{R}} = \mu_{_{F}} = \pt^{jet}$ scales~\cite{Abazov:2008hua}. Variations of PDF 
and/or scales in the jet calculations, result in differences typically of order 10\%~--~15\% for both 
measurements~\cite{Aaltonen:2008eq,Abazov:2008hua}. 

Given the agreement between the jet data and the fixed-order 
calculations\footnote{Prompt photon data at the Tevatron are also very well reproduced at large $\pt$ 
by NLO pQCD calculations~\cite{Aurenche:2006vj}.}, it is somehow difficult to conceive 
a strong disagreement in the hadron production channel since
the single high-$\pt$ charged particle spectrum is dominated by leading hadrons carrying 
out a large fraction, $\mean{z}\approx$~0.6--0.7, of the parent parton energy. Quite naively, 
 the distribution of charged hadrons above 
$\pt\sim\mean{z}\ p_T^{jet}\gtrsim30$~\GeVc\ should be also perfectly consistent with the theoretical 
predictions within the additional uncertainty introduced by the fragmentation functions which is at most
of the order $\pm$25\% as seen in Fig.~\ref{fig:nlo_uncertainties} (bottom). Let us give a rough estimate 
of the expected invariant hadron production cross section based on the jet data. At leading order accuracy and assuming 
that one partonic channel dominates jet and hadron production (e.g. $u\bar{u}\to u\bar{u}$ scattering at large $\xt$), 
the hadron production cross section Eq.~(\ref{eq:dsigma_pQCD}) is roughly given by
\begin{equation*}
{\dd^3\sigma^h\over \dd^3p}(\pt) \sim {\dd^3\sigma^{jet}\over \dd^3p}\left(\frac{\pt}{{z}}\right)\times D_u^{h^++h^-}({z}, \pt)\times \Delta z \simeq 10^{-2}\times{\dd^3\sigma^{jet}\over \dd^3p}\left(\frac{\pt}{{z}}\right)
\end{equation*}
with ${z}\simeq0.7$, the typical range $\Delta{z}\simeq 10^{-1}$ which contributes to the hadron production 
cross section, and $D_u^{h^++h^-}({z}, \pt)\simeq10^{-1}$ the $u$-quark-to-hadron FF (see e.g.~\cite{ffgenerator}). 
Using the \dzero\ jet measurement of $\dd^2\sigma^{jet}/\dd\pt\dd{y}\simeq 200$~pb/(\GeVc) at 
$\pt=100/{z}\simeq140$~\GeVc\ and $|y|<1$~\cite{Abazov:2008hua}, 
one gets for the hadron production cross section a value of $\dd^3\sigma^h/\dd^3p\simeq2\times10^{-12}$~mb/(\GeVc)$^2$ 
which is the right order of magnitude estimate expected in QCD (see the LO \pythia\ curve in Fig.~\ref{fig:pythia_cdf}).
The inconsistency between jet and the CDF large-$\pt$ hadron spectrum is also discussed in detail 
in~\cite{Cacciari:2010yd,Yoon:2010fa}.


Of course the above argument relies on the factorization assumption that large-$\pt$ hadron production 
can be expressed as a convolution of hard matrix elements with parton-to-hadron fragmentation functions. 
Should the FFs --~mostly based on fits of $e^+e^-$ data~--  be non-universal one could imagine that the 
recent CDF measurement actually reflects dramatic modifications of fragmentation functions 
in hadronic collisions. This however seems unlikely given the success of the DSS~\cite{dss} 
and AKK08~\cite{akk08} global fits of fragmentation functions which consistently use $e^+e^-$ data together 
with RHIC measurements in \pp\ collisions at $\sqrts=200$~GeV. It is in particular unclear how possible 
factorization breaking effects could enhance hadron production cross sections by up to 3 orders of 
magnitude at transverse momenta as large as $\pt=150$~\GeVc. 

There exist even more general arguments why the large-$\pt$ CDF data cannot be understood as coming from 
hadron production in perturbative QCD. As discussed in the previous section (Sect.~\ref{sec:xtscaling_tevatron}), 
the recent CDF spectrum departs from the $\xt$-scaling behaviour observed in the lower-energy data which can 
all be described assuming a scaling exponent $n\simeq 5.5$. We find on the contrary that the scaling exponent 
obtained from the comparison of the large-$\pt$ CDF measurement with the UA1 data at $\sqrts=200$~GeV is 
roughly\footnote{The precise value is difficult to obtain since the $\xt$-spectra at the two c.m. energies have a 
different slope, already indicating a non-conventional behaviour in one of the two data sets.} $n\simeq4$--4.7.  
This value is extremely close to what is expected in the conformal limit ($n=4$), 
i.e. assuming no scaling violations at all in QCD. It is in particular smaller than the exponents expected 
for jet and prompt photon production~\cite{Arleo:2009ch}, despite the fact that scaling violations are 
expected to be stronger in the hadron production channel because of the additional fragmentation process. 
What is more, the scaling exponent $n$ obtained at fixed $\xt$ reflects the $\pt$-dependence of the hard 
partonic cross section $\hat{\sigma}\sim \pt^{-n}$. Because of the fast variation of the parton densities 
with $\xt$, the $\pt$-slope, $\alpha$, of the invariant production cross section at fixed $\sqrt{s}$ is 
expected to be somehow larger than the scale dependence of the partonic cross section, i.e. $\alpha > n$. 
Surprisingly the value $\alpha$ obtained from a fit of CDF data alone above $\pt=22$~\GeVc\ is as small as 
$\alpha\simeq3.9$, that is \emph{smaller} than the combined UA1-CDF scaling exponent $n\simeq4$--4.7 (and even lower than 
the smallest scaling exponent $n=4$!). This clearly indicates that it is not possible to describe the Run-II CDF 
measurement as coming from hadron production in perturbative QCD. Hence factorization breaking effects 
in the fragmentation channel cannot be at the origin of the present discrepancy between data and NLO theory.

We conclude from this Section that the facts that (i) the NLO calculations largely fail to reproduce 
the measured single-hadron spectrum at large $\pt$ while reproducing well the single jet 
$\pt$-differential cross sections, and (ii) that the measurement violates simple phenomenological 
expectations such as $\xt$-scaling confirmed empirically in all hadronic collisions so far, 
point to a possible experimental problem in the data above $\pt\approx$~20~\GeVc\ --~or from unknown sources 
of charged particles~-- rather than from a sudden breakdown of QCD perturbation theory in the hadron production 
channel.\\


After we finished this work, other analyses appeared~\cite{Albino:2010em,Cacciari:2010yd,Yoon:2010fa,Ioffe:2010rh} 
that point out to the same discrepancy between the CDF data~\cite{Aaltonen:2009ne} and NLO calculations. 
In~\cite{Albino:2010em}, Albino, Kniehl and Kr\"amer point out the disagreement between data and NLO theory and 
question the validity of factorization theorems for large-$\pt$ hadron production, a possibility which we exclude 
(see discussion above). In~\cite{Cacciari:2010yd} it is shown on general grounds that the spectrum measured by CDF 
is inconsistent with existing Tevatron data on the inclusive jet production cross section and the distribution of 
hadrons inside jets (a similar argument is given in~\cite{Yoon:2010fa}). This observation allows the authors to exclude, 
as well, the breakdown of factorization as a possible explanation of the data. They also conclude that new physics 
scenarios explaining the CDF excess are unlikely, yet they cannot be fully eliminated. Finally, it has been claimed 
in~\cite{Ioffe:2010rh} that weak boson decays into hadrons might explain the CDF data. This possibility is however 
excluded as shown in Section~\ref{sec:pythia}, either from the detailed \pythia\ calculations or from the order of 
magnitude estimate\footnote{We believe that the calculations in~\cite{Ioffe:2010rh} is incorrect partly because of 
the use of fragmentation functions which are two orders of magnitude larger than the usual fits from $e^+e^-$ data, 
at large $z$. We also note that the Jacobian peak in the $\pt$-spectrum is located at $\sim m_W$ instead of $\sim m_W/2$. 
Finally it is difficult to conceive why the invariant cross section scales as $\sim{s}/\pt^6$ instead of conventional 
behaviour $\sim1/\pt^4$ (we thank S. Brodsky for pointing this out), which might lead to another two order of magnitude, 
$\cO{s/\pt^2}$, overestimate in~\cite{Ioffe:2010rh}.}.


\section{Inclusive charged hadron spectra at the LHC}
\label{sec:lhc}

In this last Section of the paper we present first the \incnlo\ predictions for the charged hadron 
$\pt$-differential cross sections at mid-rapidity ($|\eta|<$~1) in \pp\ collisions in the range of 
CERN LHC energies ($\sqrts$~=~0.9~--~14~TeV) including their expected theoretical uncertainties. Second, 
we discuss two interpolation methods, based on pQCD-ratios and $\xt$-scaling, that can be used to obtain 
a baseline charged hadron \pp\ spectrum at intermediate LHC energies ($\sqrts$~=~2.76, 5.5~TeV) 
needed to compare against similar measurements to be carried out in \PbPb\ collisions. 

\subsection{\incnlo\ predictions}
\label{sec:nlo_lhc}

\begin{figure}[htbp]
\centering
\epsfig{width=7.cm,file=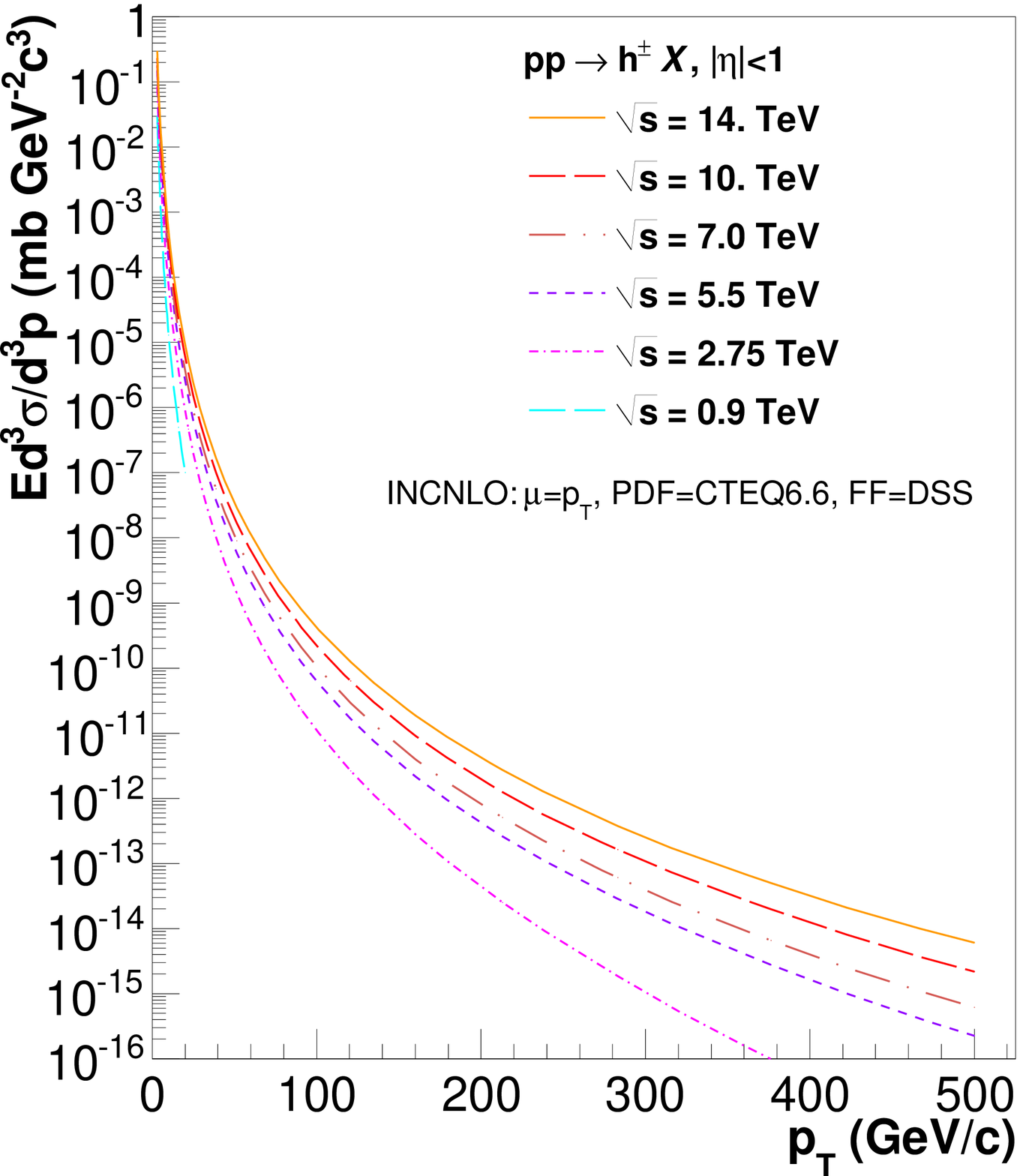}\hfill
\epsfig{width=7.cm,file=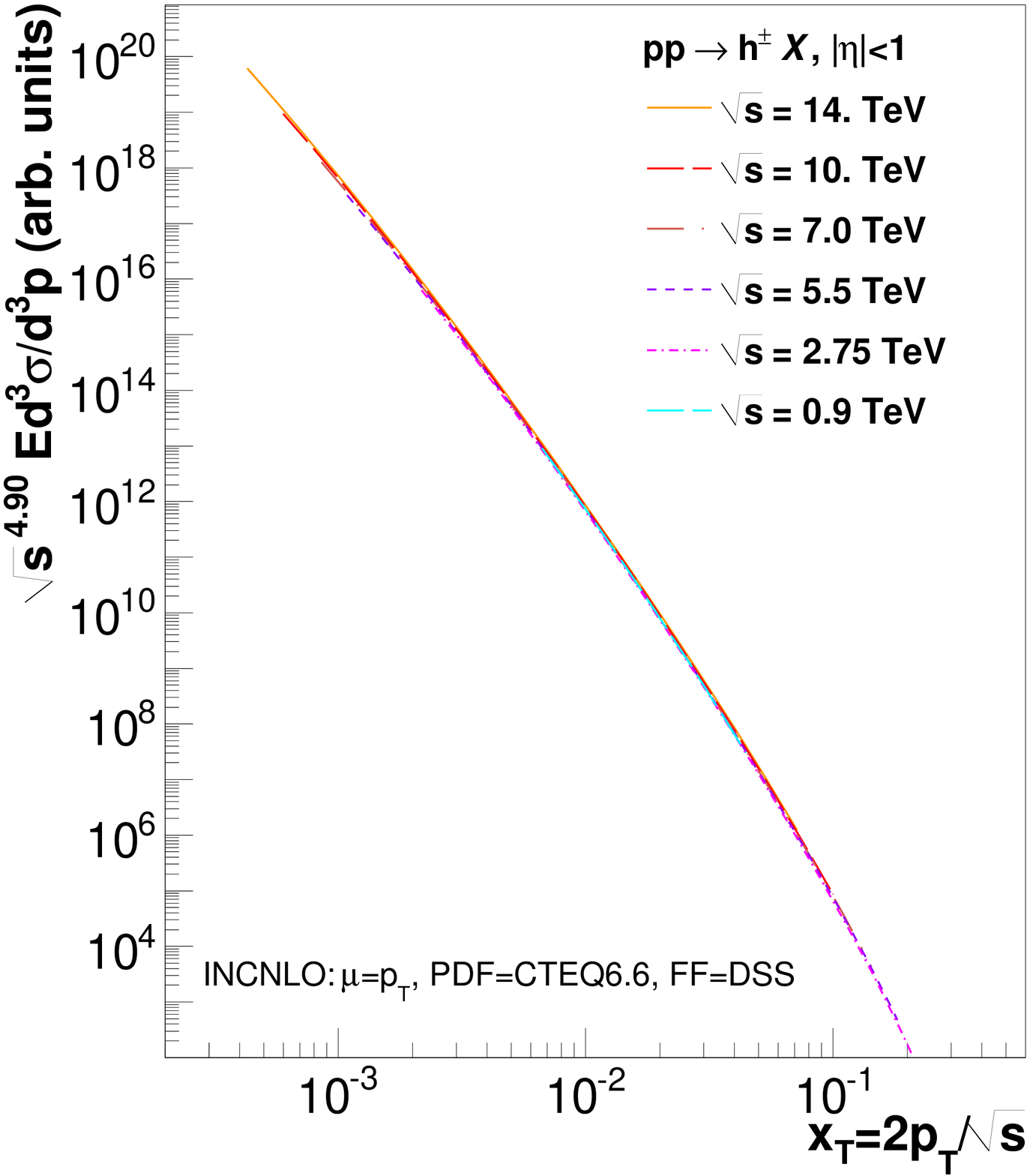}
\caption{Charged hadron spectra in \pp\ collisions at $\sqrts$~=~0.9, 2.76, 5.5, 7, 10 and 14~TeV predicted 
by NLO pQCD calculations with CTEQ6.6 parton distribution functions, DSS fragmentation functions, 
and scales set to $\mu = \pt$:~$\pt$-differential (left) and $\xt$-scaled with exponent $n=4.9$ (right).}
\label{fig:lhc_nlo}
\end{figure}

Figure~\ref{fig:lhc_nlo} (left) shows the \incnlo\ $\pt$-differential cross sections in  $p p \to h^\pm X$ 
at six different c.m. energies (expected to be) reached at the LHC at various stages of the collider programme. 
The spectra have been obtained with CTEQ6.6 PDFs, DSS FFs and theoretical scales set to  $\mu = \pt$. 
Whereas at $\pt$ below about 10~\GeVc\ all calculations converge, 
with increasing c.m. energies (and thus phase-space for hard parton-parton scatterings), 
the spectra become increasingly flatter. For example, the charged hadron yield at $\pt\approx$~100~\GeVc, 
increases by a factor of 10 between $\sqrts$~=~2.76~TeV and $\sqrts$~=~7~TeV and yet by another factor 
of 5 between the latter and the top LHC energy of 14~TeV.
As shown in Fig.~\ref{fig:lhc_nlo} (right), for this particular choice of scales/PDFs/FFs, 
a common power-law exponent of $n$~=~4.9 allows one to scale all NLO spectra in the range 
$\sqrts$~=~0.9~--~14~TeV to a universal curve, using the $\xt$ prescription given by Eq.~(\ref{eq:scaling}). 
This value is slightly smaller than what has been obtained in Section~\ref{sec:xtscaling_tevatron} 
from Tevatron and Sp$\bar{\rm p}$S measurements ($n=5.5$), indicating as expected smaller scaling 
violations at larger c.m. energy~\cite{Arleo:2009ch}.\\


\begin{figure}[htbp]
\centering
\epsfig{width=0.66\columnwidth,height=7.cm,file=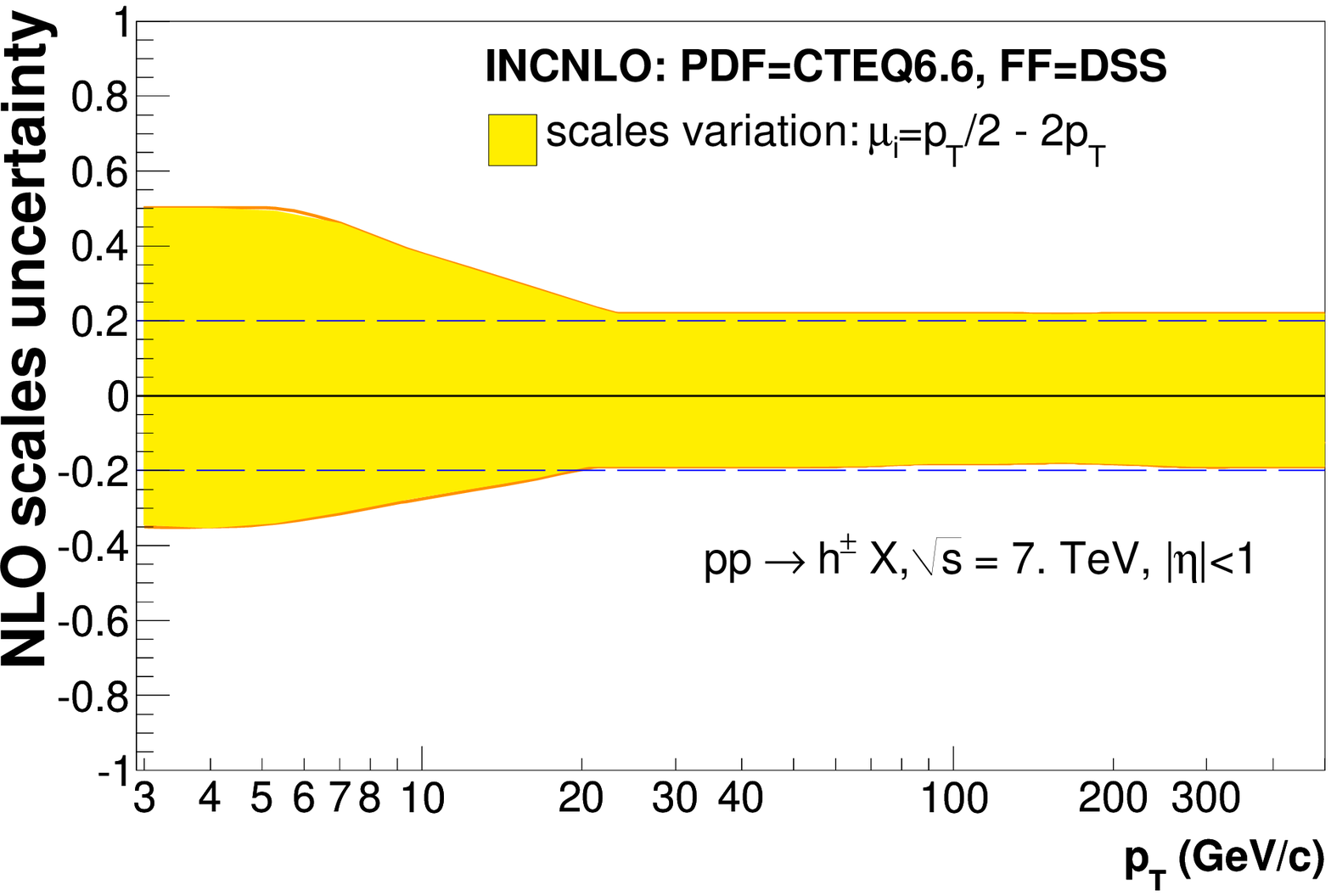}\vspace{-0.1cm}
\epsfig{width=0.66\columnwidth,height=7.cm,file=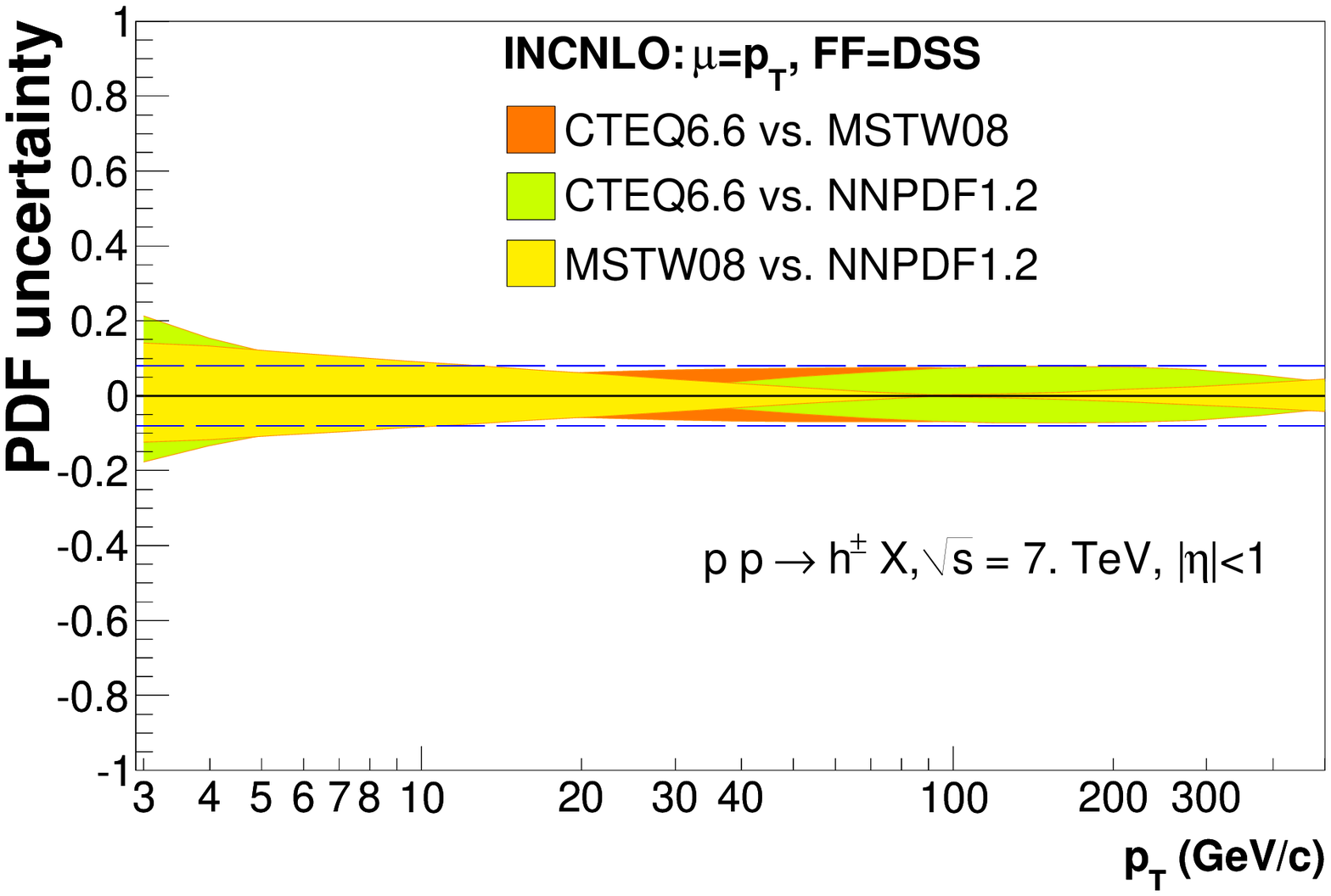}\vspace{-0.1cm}
\epsfig{width=0.66\columnwidth,height=7.cm,file=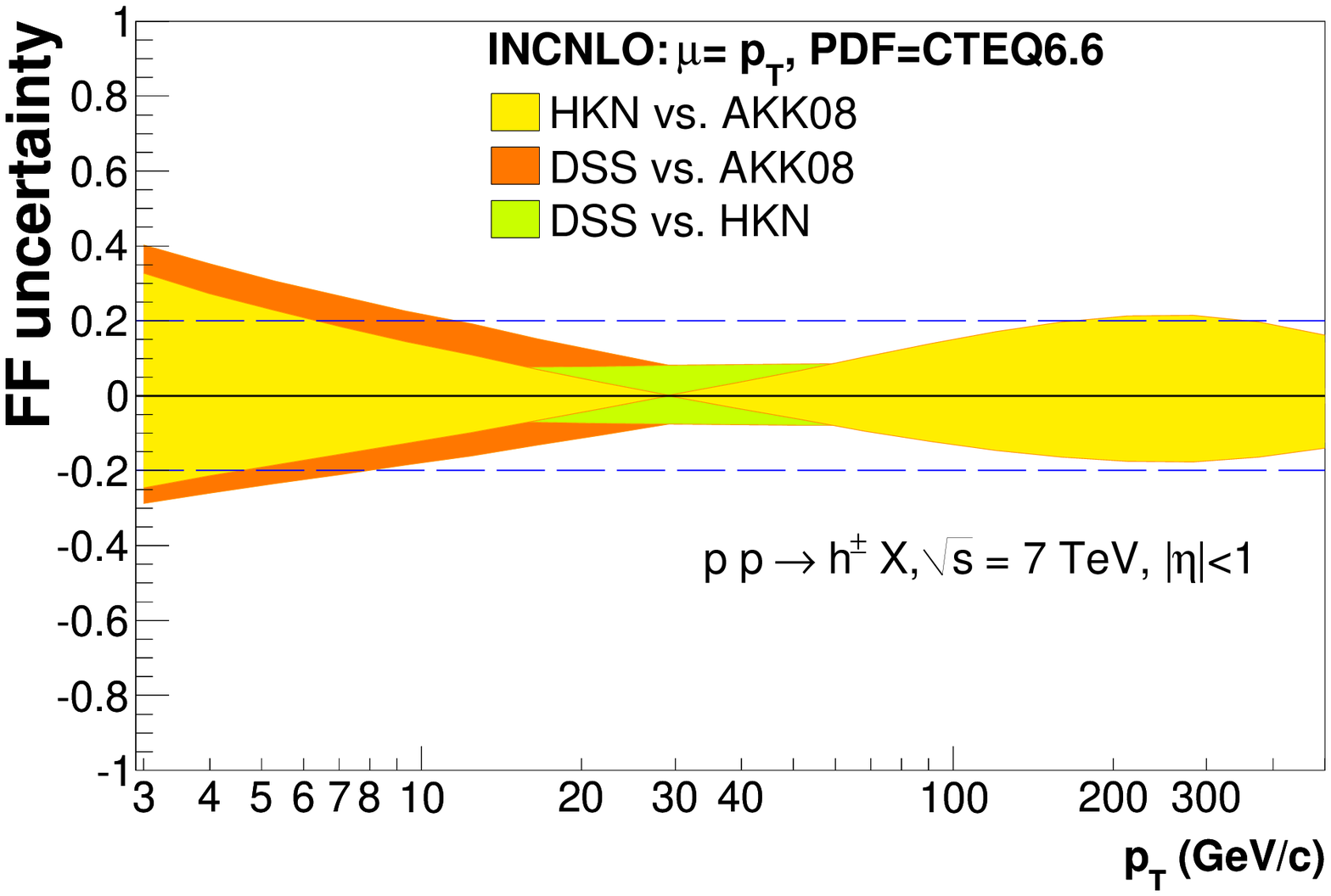}
\caption{Fractional differences between the \incnlo\ charged hadron spectra in \pp\ 
at $\sqrts$~=~7~TeV for varying scales $\mu_i$, PDF and FF.
Top: Scale uncertainty obtained for fixed PDF (CTEQ6.6) and FF (DSS) varying all three 
scales within $\mu_i~=~\pt/2 - 2\pt$ (the dashed lines indicate a $\pm$20\% uncertainty).
Middle: PDF uncertainty obtained for fixed $\mu=\pt$ and FF (DSS) with three PDFs: 
CTEQ6.6, MSTW08, NNPDF1.2 (the dashed lines indicate $\pm$8\% differences).
Bottom: FF uncertainty obtained for fixed scales ($\mu=\pt$) and PDF (CTEQ6.6) with three 
FFs: AKK08, DSS, HKNS (the dashed lines represent $\pm$20\%).}
\label{fig:lhc_nlo_uncertainties}
\end{figure}

To assess the uncertainties linked to the choice of PDFs, FFs and scales $\mu$ in the domain of
energies covered by the LHC, we have computed $pp \to h^\pm\;X$ at a fixed $\sqrts$~=~7~TeV for various 
combinations of the theoretical ingredients as done for the Tevatron prediction (see Section~\ref{sec:tevatron_incnlo}). 
Figure~\ref{fig:lhc_nlo_uncertainties} (top) shows that
the scale uncertainty is smaller ($\pm$20\% above $\pt\approx$~10~\GeVc) than found at lower 
(Tevatron) energies (see Fig.~\ref{fig:nlo_uncertainties} top).
The middle plot of Fig.~\ref{fig:lhc_nlo_uncertainties} shows that the uncertainty linked to the PDF choice
is also slightly smaller than found at Tevatron, of the order of $\pm$8\%.
Finally, the bottom panel shows that the fractional FF uncertainty 
is at most of $\pm$20\% 
above $\pt\approx$~10~\GeVc, whereas below that transverse momentum the uncertainties increase
up to $\pm$40\%. In the range $\pt\approx$~30~--~60~\GeVc\ the FF choice has uncertainties
of only 10 percent. Those results point again to a somehow smaller FF uncertainty than found at 
Tevatron (see Fig.~\ref{fig:nlo_uncertainties} bottom).
A simple quadrature addition of the fractional uncertainties linked to the scales, PDF and FF
choices results in a total theoretical uncertainty of around $\pm$35\% for the NLO 
single inclusive charged hadron spectrum in \pp\ collisions at LHC energies.\\

\subsection{Interpolation of measured charged-hadron spectra at $\sqrts$~=~5.5~TeV}

One of the assets of the successful RHIC physics program has been the ability to 
study the production of hard processes in \pp\ and nucleus-nucleus (\AaAa) collisions at the \emph{same} 
centre-of-mass energy. At the LHC, protons and ions have to travel in the same magnetic 
lattice\footnote{The magnetic rigidity is defined as $p/Z = B\,r$ for an ion with momentum 
$p$ and charge $Z$ that would have a bending radius $r$ in a magnetic field $B$.} i.e. the 
two beams are required to have the same charge-to-mass ratio $Z/A$. This limits the beam momentum of 
a given species 
to $p = 7 \,\, {\rm TeV} \times Z/A$ for the nominal 8.3~T dipole bending field. The nominal nucleon-nucleon 
c.m. energy for \PbPb\ collisions at the LHC is thus $\sqrtsnn$~=~5.5~TeV 
for lead ions with $A$~=~208 and $Z$~=~82. Since the maximum c.m. energy in the first LHC \pp\ 
runs is half of the nominal value, $\sqrts$~=~7~TeV in lieu of $\sqrts$~=~14~TeV, the first 
\PbPb\ runs are actually expected at a maximum $\sqrtsnn$~=~2.76~TeV.
In order to correctly normalize the yields measured in \PbPb\ collisions at $\sqrtsnn=2.76, 5.5$~TeV,
it will be crucial to get reliable estimates of the corresponding cross sections in \pp\ collisions 
at the same c.m. energy. 
Ideally the predictions in \pp\ collisions should take advantage of the data accumulated at the LHC 
at nearby energies and be obtained with the smallest model-dependence possible to avoid any theoretical prejudice.
In the following we present two methods for rescaling experimental \pp\
charged hadron spectra, measured at different c.m. energies than those expected for heavy-ion
collisions, based respectively on pQCD yield ratios and $\xt$-scaling.

\subsubsection{Centre-of-mass energy rescaling}

\begin{figure}[htbp]
\centering
\epsfig{width=0.7\columnwidth,file=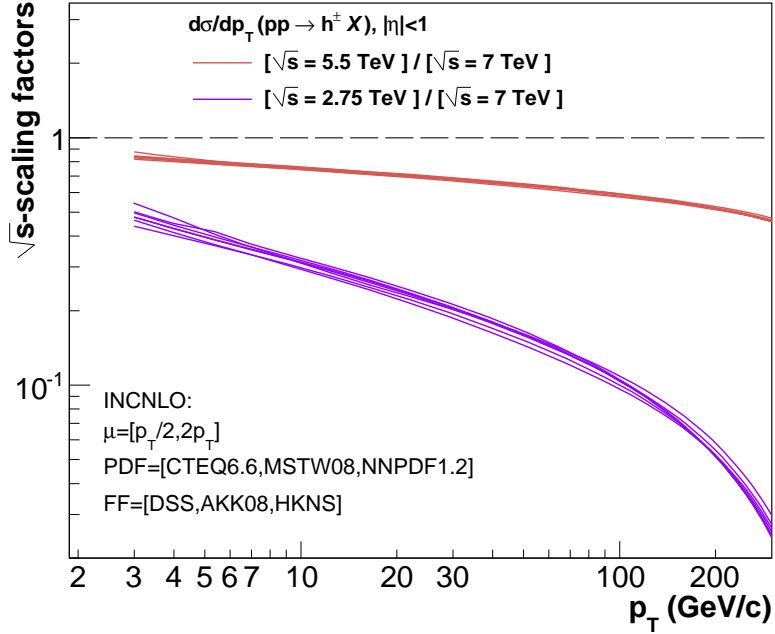}
\caption{Rescaling factors of the $\pt$-differential charged hadron cross-sections in \pp\ 
at $\sqrts$~=~7~TeV down to lower $\sqrts$~=~2.76, 5.5~TeV values, 
obtained from the ratio of the corresponding NLO calculations, Eq.~(\protect\ref{eq:rescaling}). 
The various curves show the small residual differences arising from different scale/PDF/FF choices.} 
\label{fig:rescale_comp}
\end{figure}

As seen in the previous section, at a given c.m. energy there are combined uncertainties of the order of 
$\pm$35\% in the NLO predictions for the {\it absolute} $\pt$-differential cross sections of charged hadrons
at LHC energies. 
Most of these uncertainties -- in particular the largest scale dependence -- however cancel out 
when taking {\it ratios} of the predicted perturbative yields at different, yet close, c.m. energies. 
One can, thus, rescale the \pp\ spectrum measured at a given $\sqrts$~=~$X$~TeV (say, 7~TeV) 
to a $\sqrts$~=~2.76, 5.5~TeV reference value with a simple prescription:
\begin{equation}
\frac{\dd\sigma_{\rm ref}~(\sqrts=2.76,5.5~{\rm TeV})}{\dd \pt} = 
\left(\frac{\dd\sigma_{_{\ensuremath{\it{{\rm NLO}}}}}/\dd \pt~(\sqrts=2.76,5.5~{\rm TeV})}
{\dd\sigma_{_{\ensuremath{\it{{\rm NLO}}}}}/\dd \pt~(\sqrts=X~{\rm TeV)}}\right)
\times \frac{\dd\sigma_{\rm exp}~(\sqrts=X~{\rm TeV})}{\dd \pt}.
\label{eq:rescaling}
\end{equation}
As an example, we plot in Fig.~\ref{fig:rescale_comp} the scaling factors as a function of $\pt$ 
obtained from the ratios of 2.76-TeV/7-TeV and 5.5-TeV/7-TeV pQCD yields. They have been obtained 
with up to 14 combinations of the different theoretical ingredients (scales, PDFs and FFs).
The important scale dependence (Fig.~\ref{fig:lhc_nlo_uncertainties}, top) largely cancels out 
and only residual differences arise from the slightly different parton $x$ and hadron $z$ momentum 
fractions probed at different energies (of course, the closer the c.m. energies the smaller the 
uncertainty in the yield ratios). The maximum theoretical uncertainties of the rescaling factors 
amount to a small $\pm$5\% (resp. $\pm$12\%) for $\sqrts$~=~5.5~TeV (resp. 2.76~TeV), which will be 
likely below the expected uncertainties in the 7-TeV \pp\ experimental spectrum alone.

\clearpage

\subsubsection{$\xt$-scaling}

In this last section, we suggest to use the $\xt$-scaling of particle production in high-energy scattering 
discussed in Sect.~\ref{sec:xtscaling_tevatron} in order to predict the large-$\pt$ hadron production 
cross sections in \pp\ collisions at $\sqrts=2.76$ 
and 5.5~TeV from the interpolation of Tevatron ($\sqrts=1.96$~TeV) and LHC ($\sqrts=7$~TeV) data.

Assuming that Eq.~(\ref{eq:scaling}) holds from Tevatron to LHC\footnote{We note that the Tevatron 
data are measured in $p$--$\bar{p}$ collisions unlike the \pp\ collisions at the LHC. The differences 
between both systems on unidentified hadron production at midrapidity is very small (especially, far 
away from the valence quark region, i.e. for $\xt\ll1$).}, it is straightforward to deduce the invariant 
cross section at a given $\sqrts$ and $\xt$ from previous measurements performed at Tevatron and LHC. 
Using the power-law interpolation\footnote{Note that the exponent $n$ depends in principle on  $\pt$ 
(and thus $\xt$) from the scaling violations in QCD and should approach $n=4$ in the Bjorken limit. 
However this dependence is expected to be logarithmic and can be safely neglected in the $\pt$-range being 
considered here.}, Eq.~(\ref{eq:scaling}), the invariant cross section $\sigmainv \equiv E\ d^3\sigma/d^3 p$ reads
\begin{equation}\label{eq:xs}
\sigmainv(\sqrts, \xt)=\sigmainv(\sqrts, \pt=\xt\frac{\sqrts}{2})=\sigtev\times \left[\frac{\siglhc}{\sigtev}\right]^\alpha
\end{equation}
where we define 
$\alpha \equiv \ln(\sqrts/1.96) / \ln(7/1.96)$. 
The relative uncertainty on the cross section at $\sqrts$ resulting from the power-law interpolation 
from the data at 7 and 1.96 TeV is thus simply given by
\begin{equation}\label{eq:errxs}
\frac{\delta\sigaa}{\sigaa} = \sqrt{ (1-\alpha)^2 \left( \frac{\delta\sigtev}{\sigtev} \right)^2 + \alpha^2\ \left(\frac{\delta\siglhc}{\siglhc} \right)^2}
\end{equation}
where $\delta\sigtev$ and $\delta\siglhc$ are the uncertainties of spectra measured at Tevatron and LHC, 
respectively. Let us suppose for simplicity that the experimental relative uncertainty $\delta\sigmainv/\sigmainv$
on the hadron spectrum is identical at Tevatron and LHC, thus the relative uncertainty on the interpolated 
cross section at $\sqrts$ will be $\sqrt{\alpha^2+(1-\alpha)^2}\times\delta\sigma/\sigma$, i.e. 
$0.83\ \delta\sigmainv/\sigmainv$ at both $\sqrts=2.76$ and 5.5~TeV. Therefore, if the measurements at 
Tevatron and LHC used for the interpolation are precise enough, this procedure would allow for predictions 
whose uncertainties become possibly smaller than the usual theoretical uncertainties of NLO QCD calculations. 
Note however that the uncertainty Eq.~(\ref{eq:errxs}) only reflects the propagation of errors in the 
power-law interpolation and does not account for the systematic uncertainty of the procedure currently used.

Since the invariant cross sections are compared at a given $\xt$, the $\pt$ range reached at the lower (Tevatron)
and upper (LHC) limits of the interpolation domain is crucial. The currently ``reliable'' Tevatron data extend up to 
$\pt\simeq 20$~\GeVc\ which allows for a prediction up to $\pt\simeq~30~(60)$~\GeVc\ at $\sqrts=2.76~(5.5)$~TeV. 
Conversely, the upper limit at $\sqrts$~=~2.76 (5.5)~TeV is 40\% (80\%) of the highest $\pt$ to be reached at 
$\sqrt{s}=7$~TeV.

\begin{figure}[htbp]
\centering
\epsfig{height=10.0cm,file=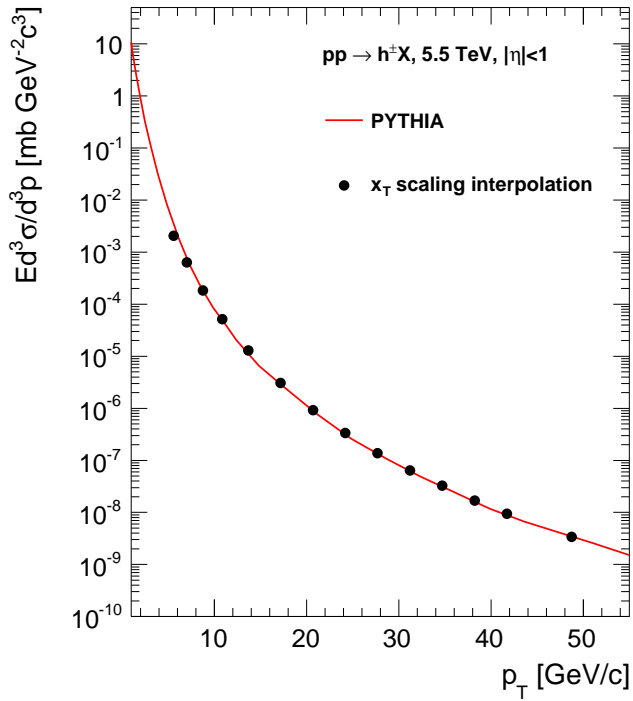} 
\vspace{-0.3cm}
\caption{
Comparison of the \pythia\ charged hadron spectrum in \pp\ collisions at $\sqrts$~=~5.5~TeV to the 
$\xt$-scaling interpolation obtained via Eq.~(\protect\ref{eq:xs}) from the corresponding \pythia\ spectra 
at $\sqrts$~=~1.96~TeV (\ppbar) and 7~TeV (\pp).}
\label{fig:pythia_xs}
\end{figure}

In order to check this procedure, the $\pt$-spectrum of mid-rapidity charged hadrons at $\sqrts=5.5$~TeV has been 
estimated from the \pythia\ spectra at $\sqrts=1.96$ and at 7~TeV using the interpolation Eq.~(\ref{eq:xs}) with 
$\alpha$~=~0.81, within the range $\xt\approx$~2$\cdot$10$^{-3}$~--~0.2. The result is plotted in 
Fig.~\ref{fig:pythia_xs} and compared to the direct \pythia\ calculation of the hadron spectrum at 
$\sqrts=5.5$~TeV. As it can be seen, the $\xt$- interpolated cross section reproduces nicely the MC result 
at $\sqrts$~=~5.5~TeV above $\pt\approx$~5~\GeVc.\\

\section{Summary}

We have compared the latest high-$\pt$ charged particle spectrum measured by CDF in proton-antiproton 
collisions at $\sqrts$~=~1.96~TeV to various perturbative QCD expectations based on next-to-leading-order 
calculations (\incnlo), parton-shower Monte Carlo (\pythia), and $\xt$-scaling respectively. 
The NLO calculations employ the latest sets of parton distribution functions (PDFs) 
and fragmentation functions (FFs). The Tevatron data can be well reproduced below $\pt\approx$~20~\GeVc\, 
for the choice of scales $\mu=2\pt$, CTEQ6.6 parton densities, and AKK08 parton-to-hadron fragmentation 
functions. Above this $\pt$ value, the CDF spectrum starts to rapidly deviate, by up to three orders of magnitude, 
from the predictions. The most important source of theoretical uncertainty, of order $\pm$30\%, 
is related to the choice of the factorization, fragmentation and normalization scales. 
The maximum uncertainties linked to the choice of the PDFs and FFs are $\pm$10\% and $\pm$25\% respectively.
A conservative quadratic addition of all these differences results in a maximum $\pm$40\% uncertainty 
in the NLO calculations which cannot by any means explain the important data--theory disagreement
above $\pt\approx$~20~\GeVc.\\

We have next determined with the \pythia\ MC the possible extra contributions of high-$\pt$ charged particles,
including leptons, coming from heavy-quark fragmentation as well as from real and virtual vector-boson 
production either single-inclusive or in association with a jet. The addition of such processes, which 
amount to about an additional ten percent of the charged particle yield above $\pt\approx$~40~\GeVc, 
does not help to reduce the large data--theory deviation. The CDF spectrum also fails to fulfill
simple $\xt$ scaling expectations which are empirically confirmed by all other high-$\pt$ hadron
spectra measured so far in \ppbar\ collisions in the range $\sqrts$~=~0.2~--~1.8~TeV. Moreover, the power-law
exponent of the CDF data above $\xt\approx$~0.02, is below the $n=4$ limit expected from simple 
dimensional arguments for pure $2\to2$ parton scattering in QCD.\\

We conclude that the fact that the NLO calculations largely fail to reproduce the measured 
CDF single hadron spectrum at large $\pt$ while simultaneously reproducing correctly the 
single jet $\pt$-differential cross sections, and that the measurement violates simple phenomenological 
expectations such as $\xt$-scaling, point to a possible experimental problem in the Tevatron data above 
$\pt\approx$~20~\GeVc\ (or to unknown sources of charged particles not considered here, 
a possibility disfavoured in~\cite{Cacciari:2010yd}).\\

The NLO predictions of charged hadron spectra at LHC energies $\sqrts=0.9$--14~TeV, have also been provided. 
Finally, we have proposed two simple interpolation procedures, based on a pQCD-rescaling and an $\xt$-scaling of 
(future) experimental \pp\ transverse momentum spectra, in order to obtain the nuclear 
modification factors of high-$\pt$ charged hadron production in nucleus-nucleus collisions at intermediate 
($\sqrtsnn$~=~2.76, 5.5~TeV) LHC energies.


\acknowledgments

FA thanks J.-P. Guillet and \'E. Pilon for discussions and the hospitality of CERN PH-TH 
department where part of this work has been completed.
DdE acknowledges support by the 7th EU Framework Programme (contract FP7-ERG-2008-235071).
A.S Yoon acknowledges support by U.S DOE grant DE-FG02-94ER40818.


\end{document}